\documentclass[twocolumn,bibyear]{aa}

\usepackage[varg]{txfonts}
\usepackage{natbib}
\bibpunct{(}{)}{;}{a}{}{,} 

\newcommand{\nc}{\newcommand}

\nc{\Msun}{\ensuremath{\mathrm{M}_\odot}}
\nc{\Rsun}{\ensuremath{\mathrm{R}_\odot}}
\nc{\cmcub}{\mbox{cm$^{-3}$}}
\newcommand{\CII}{[C \,{\sc ii}]}

\newcommand{\CI}{[C \,{\sc i}]}

\nc\micron{\mbox{$\mu$m}}
\nc{\thCO}{$^{13}$CO}
\nc{\hcoplus}{HCO$^+$}

\nc{\cmsq}{\mbox{cm$^{-2}$}}
\nc{\kms}{\mbox{km~s$^{-1}$}}
\nc{\lsun}{\ensuremath{\mathrm{L}_\odot}}

\nc{\CeiO}{C$^{18}$O}
\nc{\tex}{T$_{\rm ex}$}
\nc{\tkin}{T$_{\rm kin}$}
\nc{\rsun}{\ensuremath{\mathrm{R}_\odot}}

\usepackage[normalem]{ulem}
\usepackage{xcolor}
\definecolor{BlueF}{rgb}{0.03, 0.27, 0.49}

\definecolor{ochre}{rgb}{0.8, 0.47, 0.13}

\usepackage{graphicx}
\usepackage[varg]{txfonts}
\usepackage{natbib}
\usepackage[colorlinks=true, linkcolor=blue, citecolor=blue, filecolor=blue, urlcolor=blue]{hyperref}

\begin{document} 

\title{Weak, extended water vapor emission in the Horsehead nebula}

\author{Dariusz~C.~Lis\inst{1}
  \and Vincent~Maillard\inst{2,3}
  \and Emeric~Bron\inst{2}
  \and Franck~Le~Petit\inst{2}
  \and Javier~R.~Goicoechea\inst{3}
  \and Ducheng~Lu\inst{2}
  \and David~Teyssier\inst{4}
  }

\institute{
  Jet Propulsion Laboratory, California Institute of 
  Technology, 4800 Oak Drove Drive, Pasadena, CA 91109, USA 
  \and LUX, Observatoire de Paris, Universit\'{e} PSL, CNRS, 
  Sorbonne Universit\'{e},  92190 Meudon, France
  \and Instituto de F\'{\i}sica Fundamental (CSIC), Calle Serrano 
  121-123, 28006 Madrid, Spain
  \and European Space Agency (ESA), European Space Astronomy Centre
  (ESAC), Camino bajo del Castillo, s/n, Urbanización Villafranca del
  Castillo, Villanueva de la Cañada 28692, Madrid, Spain 
}

\date{Received 2025 August 22 / Accepted 2025 December 1}

\abstract{We analyzed archival \emph{Herschel} observations of water
  vapor emission toward the Horsehead photon dominated region (PDR),
  along with supporting ground-based and airborne observations of CO
  isotopologues and fine structure lines of ionized and atomic carbon
  to determine the distribution and abundance of water vapor in this
  low-UV illumination PDR. Water emission in the Horsehead nebula is
  very weak and, surprisingly, extends outward beyond other PDR
  tracers such as $^{12}$CO or [C\,{\sc i}]~609~$\mu$m, reaching as
  far out as [C\,{\sc ii}]~158\,$\mu$m. We model the observations
  using a newly developed PDR wrapper that takes into account the
  geometry of this region. PDR modeling of the molecular and atomic
  lines studied here provides strong constraints on the thermal
  pressure, but not on the UV illumination. Maximum model line
  intensities 
  and spatial profiles are well reproduced, except for CO
  isotopologues, where the increase on the illuminated side of the PDR
  is steeper than observed. Water vapor abundance in the model reaches
  $3.6 \times 10^{-7}$ at $A_V \sim 3$~mag. However, the ground state
  $o$-H$_2$O 557 GHz line is systematically overestimated by the
  models by at least a factor of 7 for any values of the model
  parameters. This line has a very high optical depth and the
  emergent line intensity is sensitive to radiative transfer effects
  such as line scattering by water molecules in a low-density halo
  surrounding the dense PDR and the assumed microturbulent line width.
  A more accurate model of the water surface chemistry is required. }
  

\titlerunning{Water emission in the Horsehead nebula}
\authorrunning{Lis et al.}

\maketitle 

\nolinenumbers

\section{Introduction}
\label{sec:intro}

Understanding how the Earth obtained its water and whether water-rich
earthlike planets are common in the Universe is a fundamental cosmic
origins question. Water forms efficiently in cold interstellar medium
via low-temperature ion-neutral chemistry or grain surface chemistry
\citep{vanDishoeck2013}. Gaseous water and water-ice covered
dust grains, either pristine or partially reprocessed, make their way
through the subsequent stages of star formation, including dense cores
and protostellar disks, before being incorporated into planetesimals
in forming planetary systems \citep{vanDishoeck2021}. Understanding
whether water and organics available for delivery to nascent planets
in habitable zones around stars is supplied from its initial
interstellar reservoir through a process that operates universally in
forming planetary systems is a subject of active research. The
possible links between the composition of interstellar and Solar
System materials are of great interest for understanding the Solar
System beginnings \citep{Bockelee2000,Drozdovskaya2019}.

An investigation of the ``water trail'' was one of the main science
themes of the \emph{Herschel} Space Observatory
\citep{vanDishoeck2021}. \emph{Herschel} observations demonstrated
that gas-phase water abundance is universally low, often orders of
magnitude below the canonical value of $4 \times 10^{-4}$ with respect
to H$_2$ expected if all volatile oxygen is locked in water. Water
vapor emission in molecular clouds was found to be weak and difficult
to detect away from embedded protostars. \cite{Melnick2011,
  Melnick2020} concluded that water vapor is confined primarily within
a few magnitudes of dense cloud surfaces. An exception are
photon dominated regions (PDRs), surface layers of molecular clouds
exposed to enhanced UV radiation.

\cite{Putaud2019} analyzed velocity-resolved observations of multiple
water lines in the Orion Bar, one of the best studied PDRs, where warm
chemistry with water ice desorption dominates. Using the Meudon PDR
code \citep{LePetit2006} they derived the water abundance and
ortho-para ratio in this region. They concluded that gas-phase water
arises from a region deep into the cloud, corresponding to a visual
extinction of $A_V \sim 9$. The H$_2^{16}$O fractional abundance in
this region is $\sim 2 \times 10^{-7}$ and the total water column
density is $(1.4 \pm 0.8) \times 10^{15}$ cm$^{-2}$. Contrary to
earlier suggestions, a line-of-sight averaged ortho-para ratio was
found to be consistent with a nuclear spin isomer repartition at the
temperature of the water-emitting gas, $36 \pm 2$~K.

\cite{Pilleri2012} analyzed observations of water vapor in
Mon~R2, where the associated strongly irradiated PDR could be
spatially resolved with \emph{Herschel}. They derived a low mean
abundance of $o$-H$_2$O of $\sim 10^{-8}$ relative to H$_2$, and a
higher abundance of $\sim 10^{-7}$ in the high-velocity wings detected
toward the H\,{\sc ii} region.

The Orion Bar is a dense PDR, characterized by a very high flux of UV
photons with $G_o = (3 - 5) \times 10^4$ in Habing units
\citep{Peeters2024}, and a high thermal pressure of
\mbox{$\sim (1 - 3) \times 10^8$~K\,cm$^{-3}$} \citep{Joblin2018,
  Putaud2019}. Mon~R2 is a relatively distant (830~pc) high-mass star
forming region with a complex morphology and $G_o = 5 \times 10^5$ in
its PDR. In the present manuscript we analyze observations of water
and other molecular and atomic tracers toward the Horsehead nebula, a
region characterized by a simple geometry and much lower UV
illumination with \mbox{$G_0 \sim 100$} \citep{SantaMaria2023}, more
representative of the average UV-illuminated neutral gas in the Milky
Way (e.g., \citealt{Cubick2008}). The archival data used in the
analysis are described in Section~\ref{sec:data}, the results in
Section~\ref{sec:results}, and the PDR models used to interpret the
observations are discussed in Section~\ref{sec:pdr}.
Section~\ref{sec:summary} gives a summary and conclusions.

\section{Archival data}
\label{sec:data}

The 557~GHz $1_{10}-1_{01}$ ground-state ortho-water line was observed
in the frequency-switched (FSW) mode at six positions along a right
ascension strip across the Horsehead PDR using the HIFI instrument
\citep{deGraauw2010} aboard the \emph{Herschel} Space Observatory
\citep{Pilbratt2010} as a part of the open time program
\texttt{OT\_dteyssie\_2} (\emph{Herschel} OBSID 1342203151). Spectra
were taken at right ascension offsets from --40 to
+60$^{\prime\prime}$, with 20$^{\prime\prime}$ spacing, with respect
to the center position
$\alpha_{J2000}=05^{\rm h} 40^{\rm m} 53.7^{\rm s}$,
$\delta_{J2000} = - 2^\circ 28^\prime 00^{\prime\prime}$. In addition,
FSW spectra of the same ortho-water line and the 1113~GHz $1_{11}-0_{00}$
ground-state para-water line were observed at the center position as a
part of guaranteed time program \texttt{KPGT\_vossenko\_1} (OBSIDs
1342203191 and 1342217721). The on source integration time is $\sim
30$ min per point for the 557~GHz line and 63~min for the 1113~Ghz
line.

We used in our analysis spectra obtained with the HIFI wideband
spectrometer (WBS) as a backend, which provided 1.1~MHz spectral
resolution for H and V instrumental polarizations. The two
polarization were averaged together with equal weighting. The data
downloaded from the ESA \emph{Herschel} Science
Archive\footnote{https://archives.esac.esa.int/hsa/whsa/} were reduced
with the version 14.1.0 of the HIFI pipeline. The latest values of the
HIFI beam efficiencies of 0.635 at 557 GHz and 0.59 at 1113 GHz have
been applied to convert the spectra to the main beam brightness
temperature scale.

The total observing time for the H$_2$O $1_{10}-1_{01}$ strip varied
between 20.2 and 33.1~min per position, resulting in an rms noise
level between 3.0 and 4.6~mK in a single spectrometer channel. The
integration times for the additional spectra toward the center
positions were 27.8 and 63.1~min, for ortho- and para-water lines
respectively. However, these spectra are affected by baseline ripples
and have relatively high rms noise levels of 9.3 and 18.8~mK,
respectively.

In addition to the HIFI water spectra described above we used in the
analysis ground-based observations of the Horsehead nebula from
\cite{Philipp2006}, which include $\sim 15^{\prime\prime}$
resolution single-dish images of the [C\,{\sc i}]~609 $\mu$m and CO~(4--3)
lines carried out using the CHAMP array at the Caltech Submillimeter
Observatory, as well as observations of the (2--1) transitions of the
CO isotopologues carried out with the IRAM~30-m telescope. We also use
archival observations of [C\,{\sc ii}] downloaded from the SOFIA
Science Archive at
IRSA/IPAC\footnote{https://irsa.ipac.caltech.edu/data/SOFIA/docs/data/science-archive/index.html}.

\section{Results}
\label{sec:results}

\begin{figure*} 
  \sidecaption
  \includegraphics[trim=1.5cm 3.0cm 3.5cm 1.5cm, clip=true,
  width=12cm]{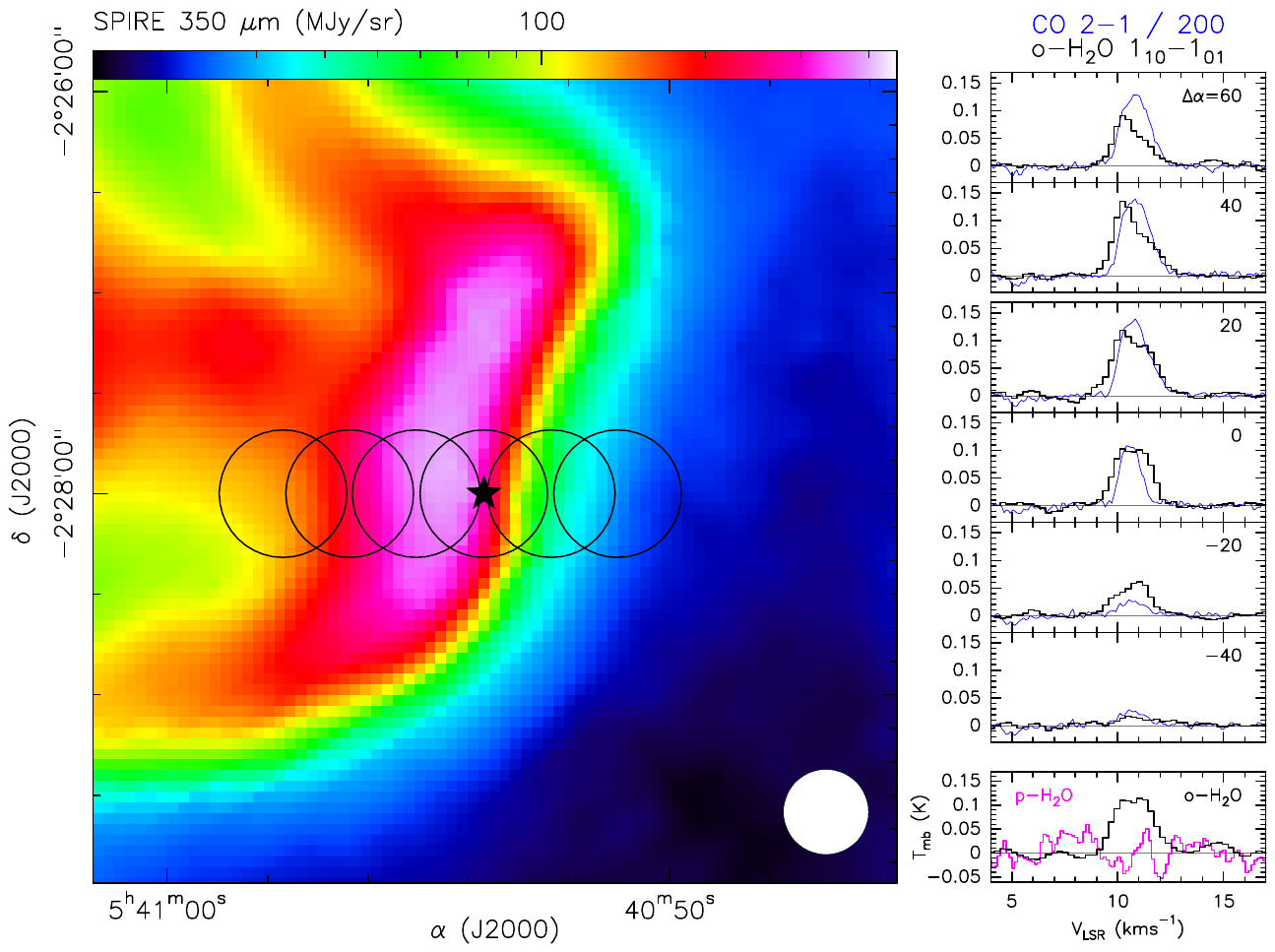} 
  \caption{(Left panel) SPIRE 350~$\mu$m image of the Horsehead PDR.
    The white circle shows the FWHM SPIRE beam
    ($25.2^{\prime\prime}$). Black circles show the locations at which
    557~GHz $o$-H$_2$O spectra were obtained, with the circle size
    corresponding to the FWHM HIFI beam ($38.1^{\prime\prime}$).
    (Right panel) Spectra of the 557~GHz $o$-H$_2$O and CO (2--1) lines
    across the PDR (black and blue curves, respectively), labeled by
    the right ascension offsets with respect to the reference
    position. The CO 2--1 spectra have been scaled down by a factor of
    200. The bottom-right panel shows spectra of the ground state
    $o-$ and $p-$H$_2$O lines at the central
    position} \label{fig:morphology} 
\end{figure*}

Figure~\ref{fig:morphology} shows the overall morphology of the
Horsehead PDR as traced by the 350~$\mu$m dust continuum emission
observed with the SPIRE instrument on \emph{Herschel}. The center
position for the HIFI water observations is marked with a black
asterisk and the black circles show locations of the HIFI beams at the
six positions along the strip. The white circle shows the FWHM SPIRE
beam at 350~$\mu$m.

The velocity-resolved HIFI water spectra of the 557 GHz along the
east-west strip across the PDR are shown as black histograms in the
top six panels on the right. The panels are labeled with the
corresponding right ascension offset with respect to the reference
position. CO~(2--1) spectra from \cite{Philipp2006} at the same
position are shown in blue. The water spectra are generally broader
than CO spectra, indicating high line opacity. In particular, the
water spectrum at the center ($\Delta\alpha = 0$) position shows a
self-absorption line profile. The bottom-right panel shows the
additional spectra of the ortho- and para-water lines at the center
position.

\begin{figure} 
   \centering           
   \includegraphics[trim=2.5cm 2.5cm 2.5cm 5.0cm, clip=true,
   width=0.95\columnwidth]{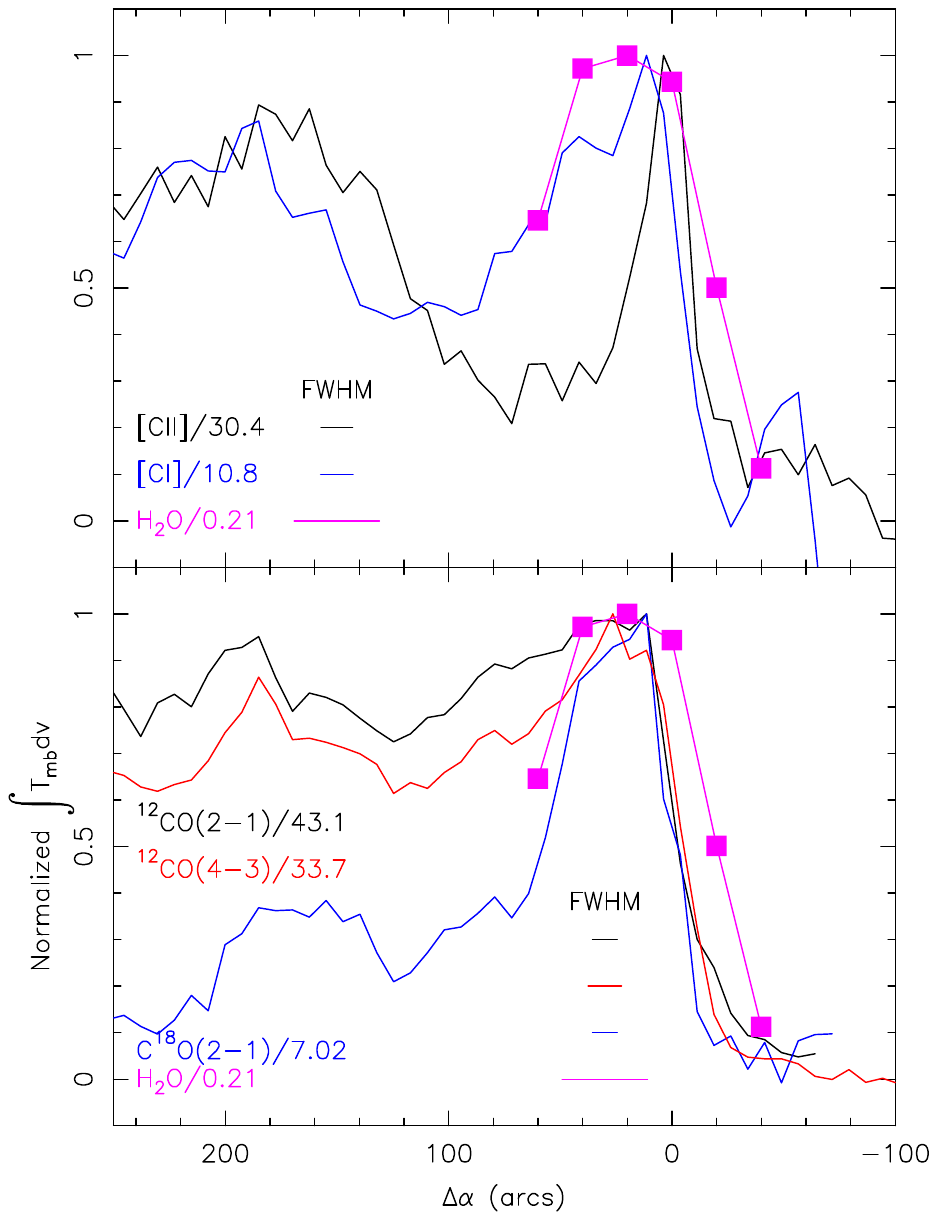}  
   \caption{Integrated line intensities of atomic and molecular
     tracers computed over 9.2--11.8~km\,s$^{-1}$ velocity range
     across the PDR as a function of right ascension offset at
     $\Delta \delta = 0$. (Upper) [C\,{\sc ii}], [C\,{\sc i}], and
     H$_2$O (black, blue, and magenta, respectively). (Lower panel)
     CO~(2--1), C$^{18}$O (2--1), CO~(4--3), and H$_2$O (black, blue,
     red, and magenta, respectively). Line intensities have been
     normalized to their maxima toward the PDR, with the corresponding
     scaling factors as labeled. The data are shown at the native
     resolution of the images, with the corresponding FWHM beam sizes
     marked for each tracer. The size of the squares
     corresponds approximately to $\pm 1 \sigma$ H$_2$O observational
     uncertainties. } \label{fig:strip}
\end{figure}

Figure~\ref{fig:strip} shows integrated line intensities of the
various atomic and molecular tracers as a function of right ascension
offset with respect to the reference position at a zero declination
offset. The upper panel shows a clear stratification between the
ionized and neutral carbon, with water emission peaking behind both
tracers. The lower panels shows the distribution of water emission as
compared to the three CO isotopologues. The water emission peaks at a
similar location C$^{18}$O, but extends further out
into the ionized region.

\begin{figure} 
   \centering
   \includegraphics[trim=2.5cm 2.5cm 2.5cm 5.0cm, clip=true,
   width=0.95\columnwidth]{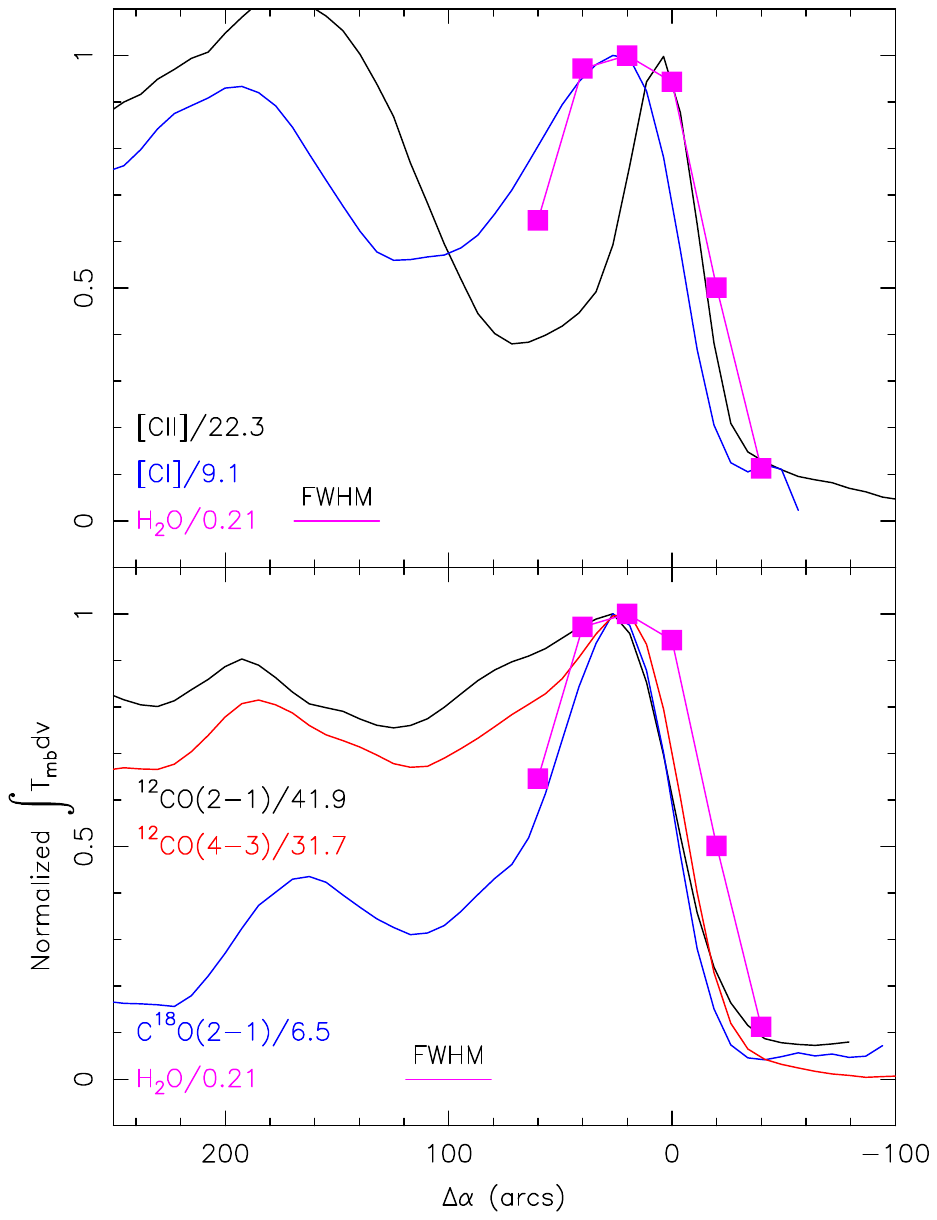}  
   \caption{Same as Figure~\ref{fig:strip}, except that all tracers
     are convolved to the $38.1^{\prime\prime}$ resolution of the HIFI
   H$_2$O spectra.} \label{fig:strip-smo}
\end{figure}

To determine whether the observed stratification may be due to the
differences in the angular resolution of the data we convolved all
images to the lowest spatial resolution of the HIFI $o$-H$_2$O data. The
resulting strips are shown in Figure~\ref{fig:strip-smo}. The same
stratification can still be seen, with water emission extending
further out into the ionized region compared to other tracers.
Integrated water line intensities between 7 and 14~km\,s$^{-1}$ at
different positions are listed in Table~\ref{tab:area} along with the
corresponding uncertainties. Maximum line intensities of the various
atomic and molecular tracers considered here in the HIFI beam are
listed in Table~\ref{tab:peak}.

\begin{table}
\begin{center}  
\caption{Water line intensities across the Horsehead PDR.} %
\label{tab:area}
\begin{tabular}{lccc}
\hline \hline 
  \rule[-3mm]{0mm}{8mm}  Line & $\Delta\alpha$   & Line intensity$^a$ & Line intensity \\
  \rule[-3mm]{0mm}{0mm}      & ($^{\prime\prime}$) & (K\,kms$^{-1}$) &
  (erg~cm$^{-2}$~s$^{-1}$~sr$^{-1}$ \\
\hline 
  $o$-H$_2$O  $1_{10}-1_{01}$ & $-40$ & $0.024 \pm 0.0036$ & $4.25 \times 10^{-9}$\\
                             & $-20$ & $0.105 \pm 0.0039$ & $1.86 \times 10^{-8}$\\
                             & 0     & $0.197 \pm 0.0053$ & $3.49 \times 10^{-8}$\\
                             & 20    & $0.210 \pm 0.0044$ & $3.72 \times 10^{-8}$\\
                             & 40    & $0.204 \pm 0.0036$ & $3.61 \times 10^{-8}$\\
                             & 60    & $0.135 \pm 0.0053$ & $2.39 \times 10^{-8}$\\
  \hline
  $o$-H$_2$O  $1_{10}-1_{01}$ & 0     & $0.222 \pm 0.012$  & $3.93 \times 10^{-8}$\\
  $p$-H$_2$O  $1_{11}-0_{00}$ & 0     & $<0.066$           & $<9.3 \times 10^{-8}$\\
\hline
\end{tabular}
\end{center}
Note: Integrated line intensities and the corresponding uncertainties
are computed over 9.2--11.8~km\,s$^{-1}$ velocity range, corresponding
to the Horsehead. The value listed for the p-H$_2$O line is a
$3 \sigma$ upper limit. $^a$ The uncertainties quoted are statistical
uncertainties computed from the rms noise in the spectra and do not
include calibration uncertainties, which should be the same for all
spectra. 
\end{table}

\begin{table*}
\begin{center}  
\caption{Maximum integrated line intensities toward the Horsehead PDR.} 
\label{tab:peak}
\begin{tabular}{lccccc}
\hline \hline 
  Tracer  & FWHM & \multicolumn{2}{c}{Native resolution} &
     \multicolumn{2}{c}{Convolved to 38.1$^{\prime\prime}$}\\
  \rule[-3mm]{0mm}{0mm}
          & ($^{\prime\prime}$)  & (K\,kms$^{-1}$) &
     (erg~cm$^{-2}$~s$^{-1}$~sr$^{-1}$)  & (K\,kms$^{-1}$) & (erg~cm$^{-2}$~s$^{-1}$~sr$^{-1}$) \\ 
\hline 
  [C\,{\sc ii}]      &  14.1 & 30.4 & $2.14 \times 10^{-4}$ & 22.3 & $1.57 \times 10^{-4}$ \\\relax 
  [C\,{\sc i}]       &  14.5 & 10.8 & $1.32 \times 10^{-6}$ & 9.1 & $1.11 \times 10^{-6}$ \\
  $^{12}$CO (2--1)    &  11.0 & 43.1 & $5.41 \times 10^{-7}$ & 41.9 & $5.26 \times 10^{-7}$ \\
  C$^{18}$O (2--1)    &  11.0 & 7.02 & $7.61 \times 10^{-8}$ & 6.5 & $7.05 \times 10^{-8}$ \\
  $^{12}$CO (4--3)    &  15.0 & 33.7 & $3.38 \times 10^{-6}$ & 31.7 & $3.18 \times 10^{-6}$\\
  $o$-H$_2$O  $1_{10}-1_{01}$ & 38.1 & 0.21 & $3.72 \times 10^{-8}$ & 0.21 & $3.72 \times 10^{-8}$ \\
\hline
\end{tabular}
\end{center}
Note: Maximum values of the integrated line intensities at the native
resolution of the data and in a 38.1$^{\prime\prime}$ beam,
computed over 9.2--11.8~km\,s$^{-1}$ velocity range.
\end{table*}

\section{Pseudo 2-D PDR modelling of the Horsehead nebula}
\label{sec:pdr}

\subsection{Previous approaches to model the Horsehead geometry with
  1D plane-parallel codes} 

The Horsehead nebula has long served as a benchmark for the study of
PDRs, owing to its well-defined, nearly edge-on geometry and wealth of
multi-wavelength observational data (e.g., \citealt{Pety2005}).
Despite its apparent spatial simplicity, modeling the Horsehead poses
specific challenges. Most PDR models, including those using the
Meudon PDR code, adopt a 1D plane-parallel slab geometry in which the
radiation field and physicochemical gradients vary the
depth into the cloud, as the UV photons are gradually attenuated by
dust. However, in the case of the Horsehead, the PDR is
viewed nearly edge-on, meaning that the observer's line of sight is
perpendicular to the primary gradient direction in the model. As a
result, direct comparisons between model predictions (computed along
the depth axis) and observational quantities (integrated along the
line of sight) require nontrivial adaptations, typically involving
assumptions about the effective length of the PDR along the line of
sight or geometrical corrections.
        
\cite{Habart2005}, using SOFI observations of the H$_2$ 1--0 S(1) line
at the NTT, modeled the Horsehead nebula as an edge-on, semi-infinite
plane-parallel slab illuminated by an external radiation field and
exhibiting a steep density gradient. This structure corresponded to an
isobaric PDR with a thermal pressure of $4 \times 10^6$~K~cm$^{-3}$.
The computed emissivity was then scaled by the PDR length along the
line of sight, assumed to be at most equal to the filament's extent in
the plane of the sky, i.e., approximately 0.1~pc.

\cite{Goicoechea2006} first used the output of Meudon PDR models as
input for non-LTE radiative transfer models of CS and C$^{18}$O
emission lines, adapted to the edge-on geometry of the PDR. Similarly,
\cite{Pabst2017} used the predictions of a face-on PDR model to
calculate the expected [C \textsc{ii}] 158~$\mu$m line emission. Both
approaches compute the line-of-sight radiative transfer assuming 1D
plane-parallel slabs.
 
\cite{Guzman2011} also employed Meudon PDR models with a steep density
gradient to interpret H$_2$CO transition lines observed with the
IRAM-30m telescope, which were converted into chemical abundances.
These derived abundances were compared to model predictions with and
without surface chemistry, at two positions in the cloud referred to
as the "core" and the "PDR". A similar methodology (direct comparison
between observed abundances and Meudon PDR model outputs) was applied
in the interpretation of IRAM-30m observations of CF$^+$,
l-C$_3$H$^+$, CH$_3$OH, and HCO in \cite{Pety2012, Guzman2012,
  Guzman2013}.

Using sub-arcsecond angular resolution ALMA observations of the CO
$J=3$–2 line, \cite{Hernandez2023} constrained the thermal pressure in
the Horsehead nebula to the range
$(3.7 - 9.2) \times 10^6$~K~cm$^{-3}$. This constraint was derived by
comparing the observed spatial offsets between the C/CO and H/H$_2$
transitions with respect to the H/H$^+$ transition, to those predicted
by 1D Meudon PDR models. Only models within the specified thermal
pressure range produced transition separations of the same order of
magnitude as observed.

\subsection{The PDR wrapper approach}

Here, we also use Version 7 of the Meudon PDR
code\footnote{https://pdr.obspm.fr}, but to compare the outputs of the
code with spatially resolved observations of edge-on structures such
as the Horsehead nebula, we developed a dedicated post-processing
tool: the \textit{PDR wrapper}. This tool maps the one-dimensional
Meudon PDR model onto a two-dimensional plane defined by the lines of
sight and the direction of radiation propagation, under the assumption
of a simplified, shell-like geometry characterized by a fixed radius
of curvature.

\begin{figure*}
\centering   
\includegraphics[width=0.9\textwidth]{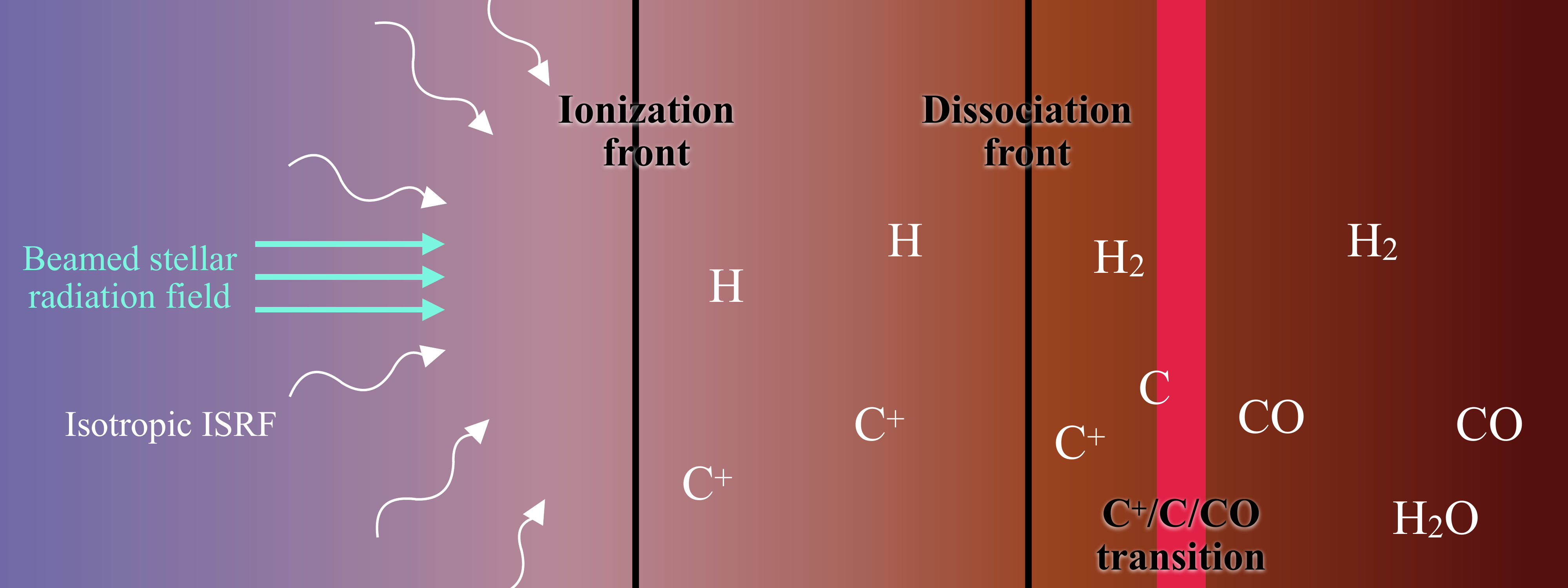}
\caption{Illustration of the geometry of one-dimensional PDR models
  made with the Meudon PDR code. We see the two-component radiation
  field composed of a beamed part representing the stellar radiation
  field, and an isotropic part representing the ISRF. This
  configuration yields the typical chemical stratification with the
  ionization front denoting the beginning of the PDR, followed by the
  dissociation front, or H/H$_2$ transition, and the C$^+$/C/CO
  transition deeper in the cloud. H$_2$O appears even deeper.} 
\label{fig:Meudon_edge-on_geometry}
\end{figure*}

Figure~\ref{fig:Meudon_edge-on_geometry} depicts a typical 1D PDR
model geometry as used by the Meudon PDR code, with a two-component
radiation field composed of a perpendicular beamed stellar radiation
field from a nearby star and an isotropic mean interstellar radiation
field (ISRF). We also see the chemical stratification, with the
ionization front denoting the beginning of the PDR, followed by the
dissociation front (H/H$_2$ transition) separating the atomic and
molecular regions, and the C$^+$/C/CO transition taking place deeper
in the cloud, followed by the appearance of H$_2$O molecules.

\begin{figure*}
\centering   
\includegraphics[width=0.9\textwidth]{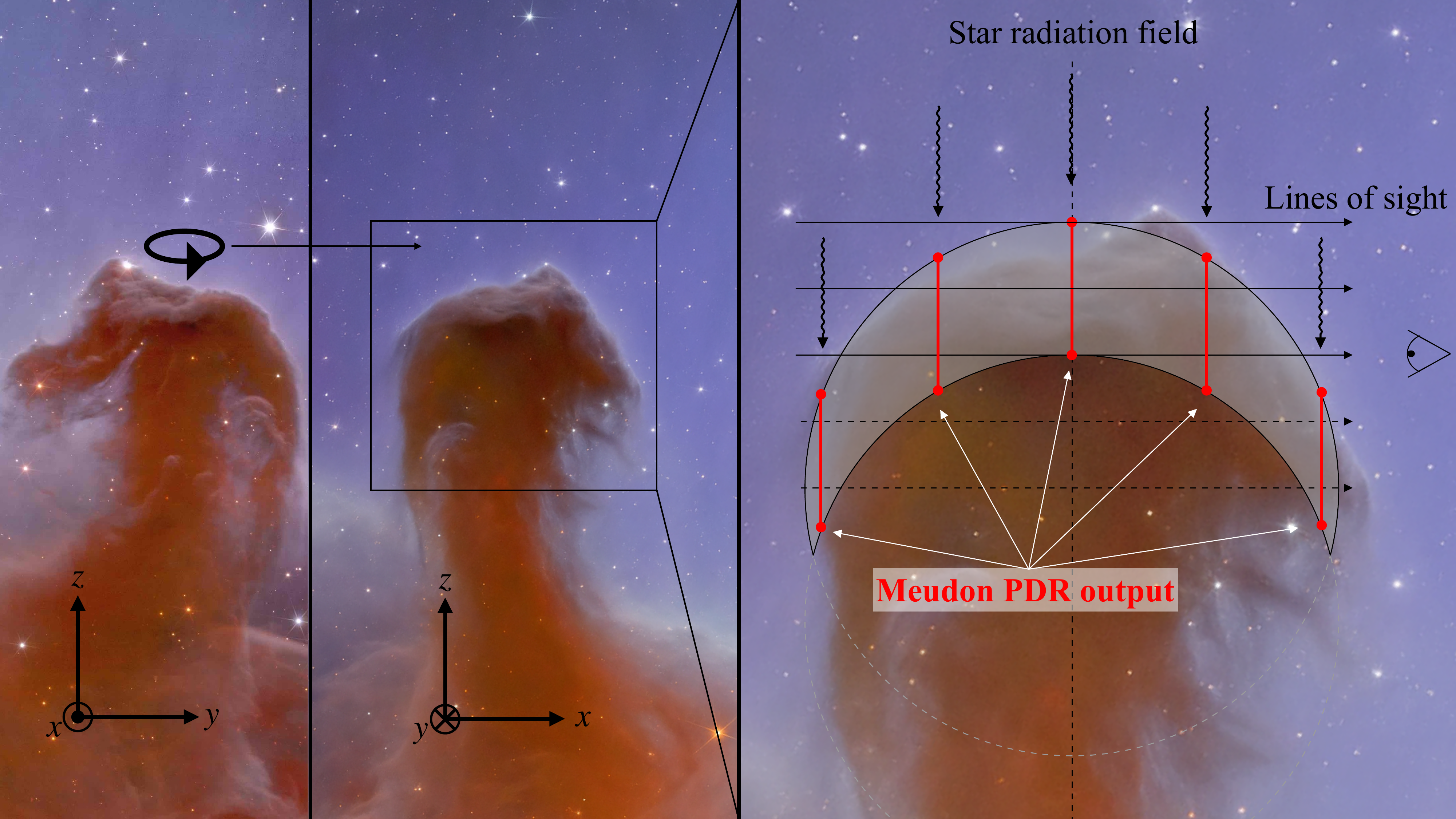} 
\caption{Illustration of the geometry of the PDR wrapper. (Left panel)
  Image of the Horsehead nebula as seen by the Euclid space telescope
  (Credits: ESA/Euclid/Euclid Consortium/NASA, image processing by
  J.-C. Cuillandre (CEA Paris-Saclay), G. Anselmi). Notice the
  reference frame, with the $x-$axis corresponding to the direction of
  the lines of sight. We rotate the nebula so that the $x-$axis points
  to the right. (Middle panel) Artistic interpretation of the rotation
  of the Horsehead nebula, generated using OpenAI’s DALL·E model based
  on the first panel image. (Right panel) Geometry of the wrapping 2D
  model, with the star radiation field coming from the top, and
  examples of Meudon PDR model output over-plotted as red vertical
  lines. Each of these lines correspond to a PDR model as represented
  on Fig. \ref{fig:Meudon_edge-on_geometry}. Notice that we use the
  same model for each position along the cut of the cloud. The
  multiple lines of sight correspond to different pixels along a
  vertical one-dimensional cut in the Horsehead nebula. We emphasize
  that only the lines of sight shown as solid lines are
  correctly accounted for by the model, as the central section of the
  nebula outside of the crescent-shaped area is not included in the
  computations. The geometry is assumed circular, with a radius
  of curvature that remains to be determined.}
\label{fig:Wrapper_geometry}
\end{figure*}

An illustration of the wrapper's conceptual geometry is provided in
Figure~\ref{fig:Wrapper_geometry}. The left panel shows a real image
of the Horsehead nebula, while the middle and right panels offer a
rotated side-view artistic rendition of the cloud. In this  
representation, the stellar radiation field is incident from the top,
while the lines of sight toward the observer extend horizontally to
the right. The total set of lines of sight corresponds to a vertical
column of pixels in the original image, effectively forming a cut
across the nebula.

The shell-like structure superimposed in the right panel models the
curved PDR surface, along which the output of a single 1D Meudon PDR
model is mapped. This 1D structure is effectively "swept" along the
arc, assuming a constant radius of curvature. Examples of such 1D
models are drawn as red vertical lines on the figure.

A run of the Meudon PDR code produces a wide range of outputs. Among
these, we make use of the density profiles of the chemical species
of interest, as well as the profiles of their
populations in individual quantum levels. Note that by using
perpendicular illumination in the PDR code to simulate irradiation by
the illuminating star, each point on the surface of the resulting
sphere is illuminated perpendicularly. This approach provides a more
realistic representation of a stellar-illuminated surface than would
do a 1D spherical PDR model, in which the illumination is assumed to
be isotropic.

\begin{figure*}
\centering
\includegraphics[width=0.3\textwidth]{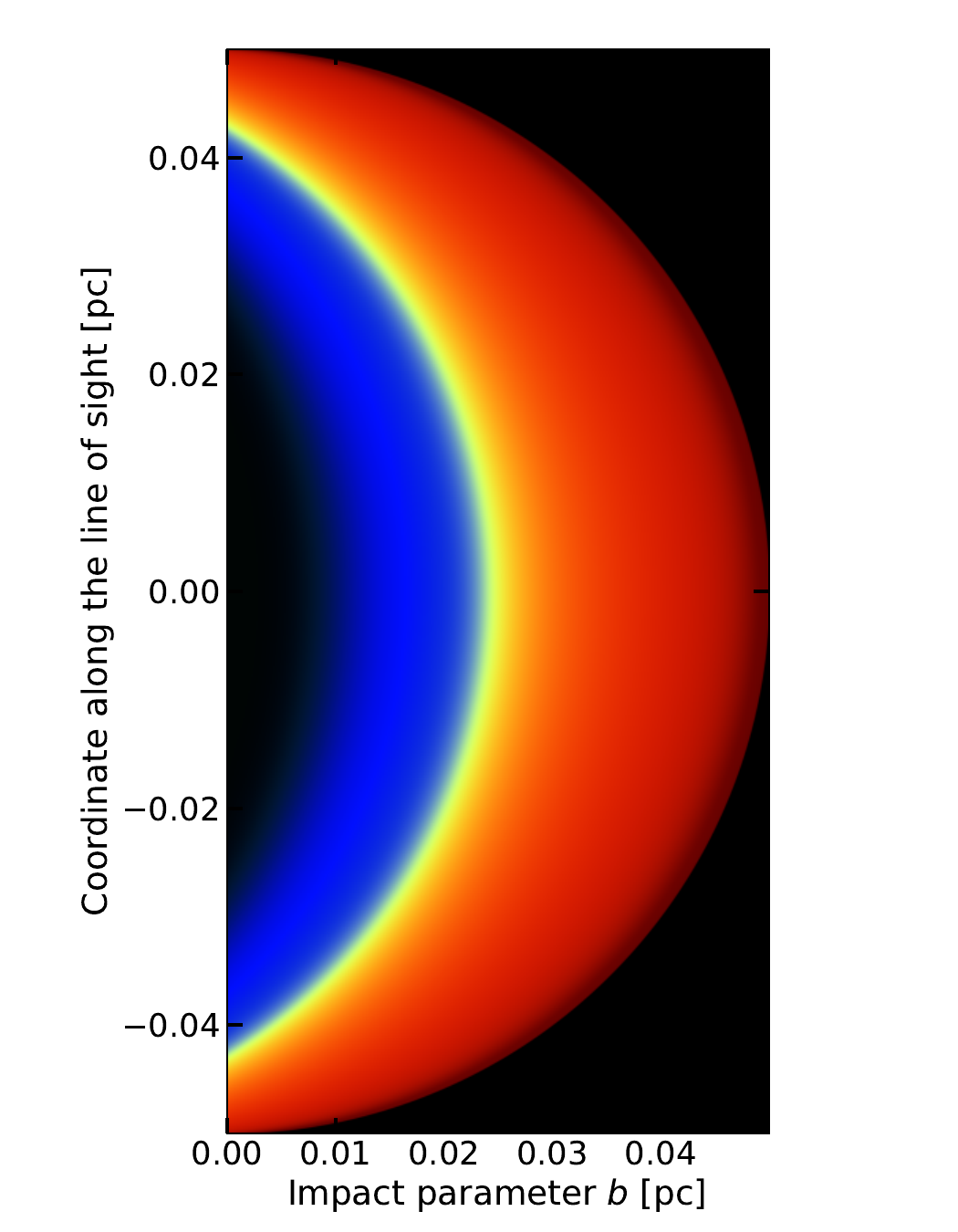} 
\includegraphics[width=0.3\textwidth]{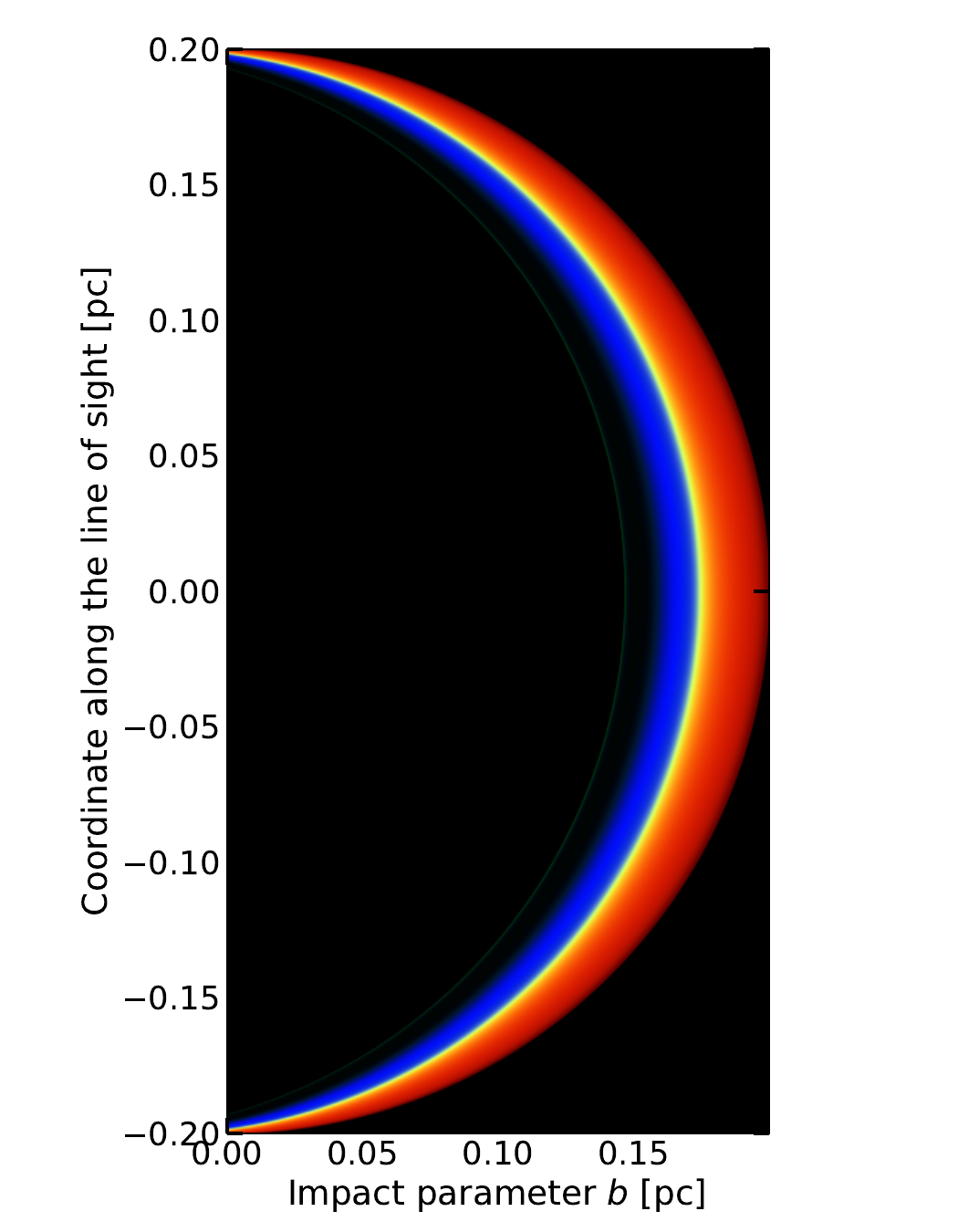}
\includegraphics[width=0.3\textwidth]{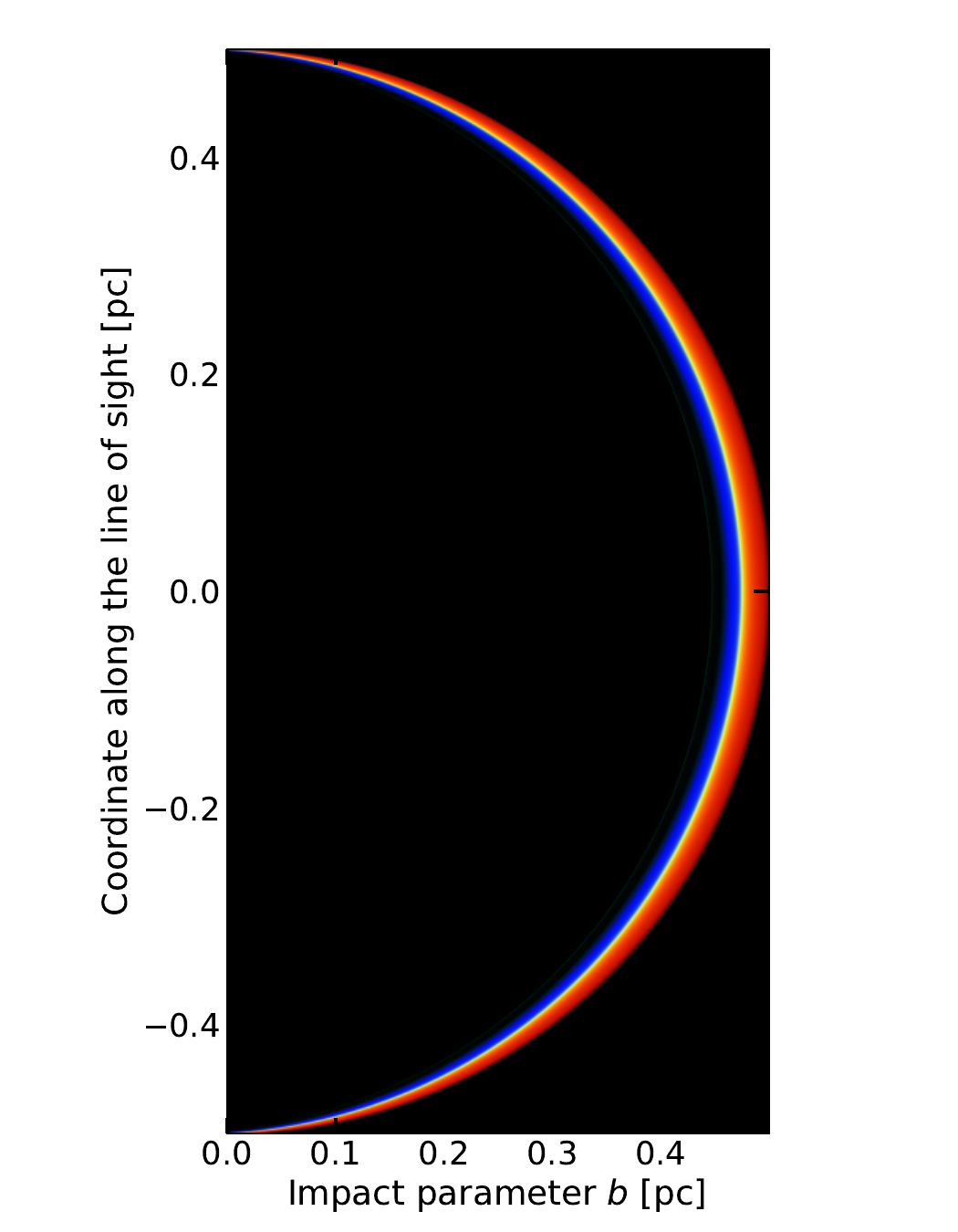}
\caption{Illustration of the wrapper geometry, showing RGB composite
  maps of carbon-bearing species for different radii of curvature. The
  color encodes the local abundances as follows: red for ionized
  carbon (C$^+$), green for neutral atomic carbon (C), and blue for
  carbon monoxide (CO). Areas where these species coexist appear as
  color blends (e.g., cyan for C and CO, yellow for C$^+$ and C, white
  for equal contributions). The left panel corresponds to a curvature
  radius of $R_{\text{C}} = 0.05$ pc, the middle to $R_{\text{C}} =
  0.2$ pc, and the right to $R_{\text{C}} = 0.5$ pc. These maps
  highlight the local spherical curvature in the plane of the sky,
  with the observer located at the bottom of each panel and the
  stellar radiation field entering from the right. Radiative transfer
  is computed along vertical lines from top to bottom in this
  configuration. The abundances originate from a single Meudon PDR
  model ($P_{\mathrm{th}} = 4 \times 10^6$~K\,cm$^{-3}$, $G_0 = 100$).
  The usual edge-on representation of abundances of this model can be
  found in Figure \ref{fig:edge-on}.} 
\label{fig:Wrapper_mapping}
\end{figure*}

Once the level populations are known throughout this two-dimensional
structure, we compute the line intensities as seen by an observer
along many perpendicular lines of sight. This allows us to reconstruct
the spatial variation of the line intensities from the ionization
front to the core of the cloud. The line intensities are determined by
solving the radiative transfer equation along each line of sight,
using the level populations interpolated from the Meudon PDR model,
integrating from the far side of the cloud to the edge of the PDR on
the observer's side. See Appendix~\ref{app:wrapper} for more details.

Lines of sight shown as solid arrows in
Figure~\ref{fig:Wrapper_geometry} intersect only physically meaningful
regions of the cloud where the PDR model predicts densities and level
populations. In contrast, dashed lines cross into central regions not
covered by the model and are excluded from further analysis.

The wrapper reconstructs the spatial emission profiles as they would
appear when observing a locally curved PDR edge-on, accounting for
both curvature effects and radiative transfer along the lines of
sight. This method enables the generation of synthetic spatial
emission profiles while avoiding the computational cost of full
two-dimensional radiative transfer, which can be prohibitive for large
parameter-space explorations while maintaining a very high level of
process modelling.

Examples illustrating the wrapper geometry are shown in
Figure~\ref{fig:Wrapper_mapping}, where a single Meudon PDR model
($P_{\mathrm{th}} = 4 \times 10^6$~K\,cm$^{-3}$, $G_0 = 100$ Habing
units, \citealt{Habing1968}) is projected onto shells with three
different radii of curvature. The RGB composite maps simultaneously
represent the abundances of the three main carbon-bearing species: red
for ionized carbon (C$^+$), green for atomic carbon (C), and blue for
carbon monoxide (CO). This color-coding highlights the chemical
stratification from the ionized outer layers to the molecular
interior, with mixed regions appearing as intermediate colors such as
yellow, cyan, or white. The observer is located at the bottom of each
map, and the stellar radiation field enters from the right. These
examples visualize the transition from ionized to atomic to molecular
carbon, as well as the local spherical curvature along the line of
sight. In this representation, radiative transfer is computed from top
to bottom.

\subsection{Model grid and physical ingredients}
\label{sec:model_grid}
        
As discussed in Section~\ref{sec:H2O_Wrapper}, reproducing the H$_2$O
emission line proved challenging. For this reason, we separate the
analysis into two parts: the reproduction of the C$^+$/C/CO
transitions, based on both line intensities and spatial profiles, is
presented here, while the modelling of the H$_2$O line and its spatial
distribution is treated independently.

For the first part of the analysis, we constructed a grid of isobaric
Meudon PDR models coupled with the wrapper tool. The explored
parameter space includes:
\begin{itemize}
\item Thermal pressures $P_{\mathrm{th}} = 0.5$, 1, 2, 3, 4, 5, 6, 8,
  and $10 \times 10^6$~K\,cm$^{-3}$;
\item Curvature radii $R_{\mathrm{C}} = 0.02$, 0.05, 0.1, 0.2, 0.5,
  and 1.0~pc. 
\end{itemize}

A standard value for the interstellar radiation field of $G_0 = 100$
Habing units is used for the models, as the line intensities of
[C\,\textsc{ii}]~158~$\mu$m, [C\,\textsc{i}]~609~$\mu$m, and low-$J$
CO transitions depend only weakly on $G_{0}$ in low-excitation PDRs
such as the Horsehead nebula (see e.g. \citealt{Kaufman1999}). This
choice of $G_0$ is motivated by multiple studies of the Horsehead
nebula (e.g., \citealt{Abergel2003,Habart2005,Goicoechea2009,
  SantaMaria2023,Zannese2025}) and the
spectral type of the illuminating star, $\sigma$~Ori, O 9.5 V at a
projected distance of 3.5~pc. We further discuss the weak $G_0$
dependence in Appendix \ref{app:Appendix_Indep_to_G0}.

The models incorporate an exact radiative transfer treatment for H$_2$
lines. This means that H$_2$ self-shielding is calculated precisely,
properly accounting for the overlap of UV pumping lines from both H
and H$_2$. This formalism also ensures that the model accurately
considers the shielding of the CO UV lines by the H$_2$ Lyman and
Werner transitions. 

The chemical network considers 220 species and more than 3\,300
reactions including a surface chemistry network for the chemistry of
carbon and oxygen and the formation of mantles of H$_2$O on the
grains. The radiation field is modeled using a synthetic spectra from
the POLLUX database \citep{PolluxDataBase} to simulate the
$\sigma$-Ori stellar spectrum. The selected parameters are
$T_{\textrm{eff}}$ = 33\,000 K, log(g) = 4.26 and R = 5.26 R$_\odot$.
Turbulent broadening is included with a velocity dispersion
$\sigma_{\mathrm{turb}} = 0.38$~km\,s$^{-1}$, following
\citet{Hernandez2023}, derived from ALMA observations of HCO$^+$ in
the Horsehead nebula. The impact of this choice on the predicted line
intensities is discussed in Sect.~\ref{sec:H2O_Wrapper}.

\subsection{Model fitting procedure and main results}\label{sec:model}

The synthetic emission profiles produced by the wrapper were compared
to observational data using a $\chi^2$ minimization method implemented
in the \texttt{pyismtools} Python suite. The fitting simultaneously
considered both the maximum intensities and, for the first time, the
spatial separation between the emission peaks along the PDR. For the
minimization, we did not use the raw output of the wrapping procedure,
but rather the profiles convolved to match the native angular
resolution of each tracer image.

\begin{figure*}
\sidecaption
\includegraphics[width=12cm]{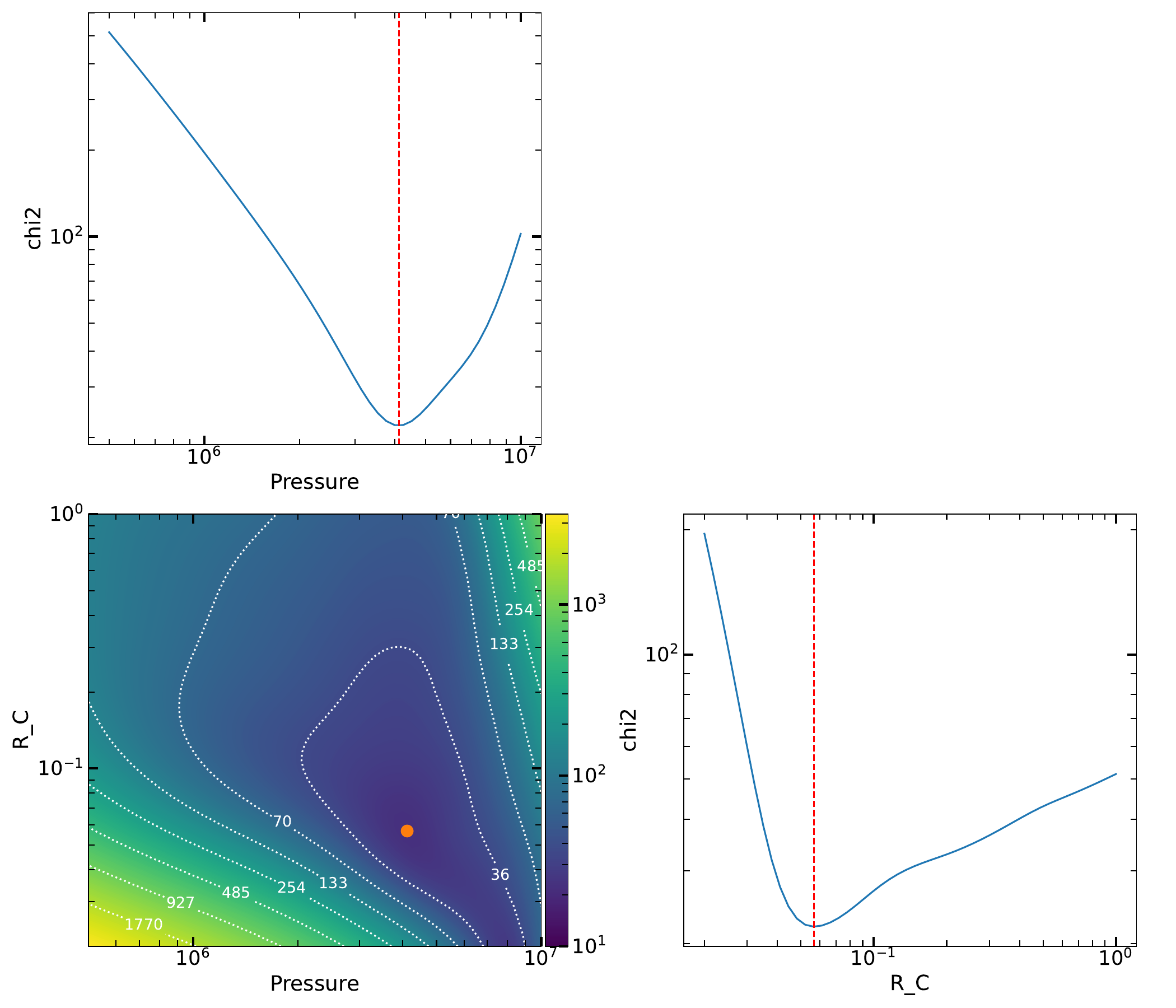} 
\caption{
  $\chi^2$ map obtained from the minimization procedure using maximum
  intensities and spatial separations. The best-fit model,
  corresponding to $P_{\mathrm{th}} = 4.1 \times 10^6$~K\,cm$^{-3}$,
  and $R_{\mathrm{C}} = 0.057$~pc, is denoted by the orange
  circle on the map. The constraint on $R_{\mathrm{C}}$
  remains weak, with the $\chi^2$ surface relatively flat in a wide valley.
  }
\label{fig:chi2_map}
\end{figure*}

The minimum $\chi^2$ value is obtained for
$P_{\mathrm{th}} = 4.1 \times 10^6$~K\,cm$^{-3}$ and
$R_{\mathrm{C}} = 0.057$~pc, 
being consistent with previously published,
with $R_{\mathrm{C}}$ being around half the typical 0.1 pc width usually
assumed in plane-parallel models of the Horsehead PDR.

A comparison between the best model and the observations is
shown in Fig.~\ref{fig:best_model}. The bottom panel displays
the maximum intensities of the emission lines, demonstrating excellent
agreement. Model predictions are plotted as empty circles with 40\%
error bars, accounting for geometrical uncertainties and the
complexity of the parameter space.
In the top panels, we compare the \textit{full} spatial emission
profiles from our model to the observed profiles, even though the
minimization relied solely on maximum intensities and separations
between various tracers. The left panel shows the raw model profiles
produced by the Wrapper, while the right panel presents the same
profiles convolved to the native angular resolution of the respective
tracers, as used in the minimization. As stated before, we emphasize
that our modeling focuses solely on reproducing the initial rise in
intensity for each emission line, since the geometrical model is
physically consistent only within the computed width of the PDR.
        
Notably, even though only maximum positions were used for fitting, the
complete spatial profiles are well reproduced. An exception is
observed for the CO lines and their isotopologues, for which the
observed rise in intensity near the illuminated edge is smoother than
predicted. This discrepancy remains unexplained and may be due to
convolution of multiple CO fronts (eg., \citealt{Hernandez2023}).

\subsection{H$_2$O emission}
\label{sec:H2O_Wrapper}

\begin{figure*}[h!]
\centering   
\includegraphics[width=0.45\textwidth]{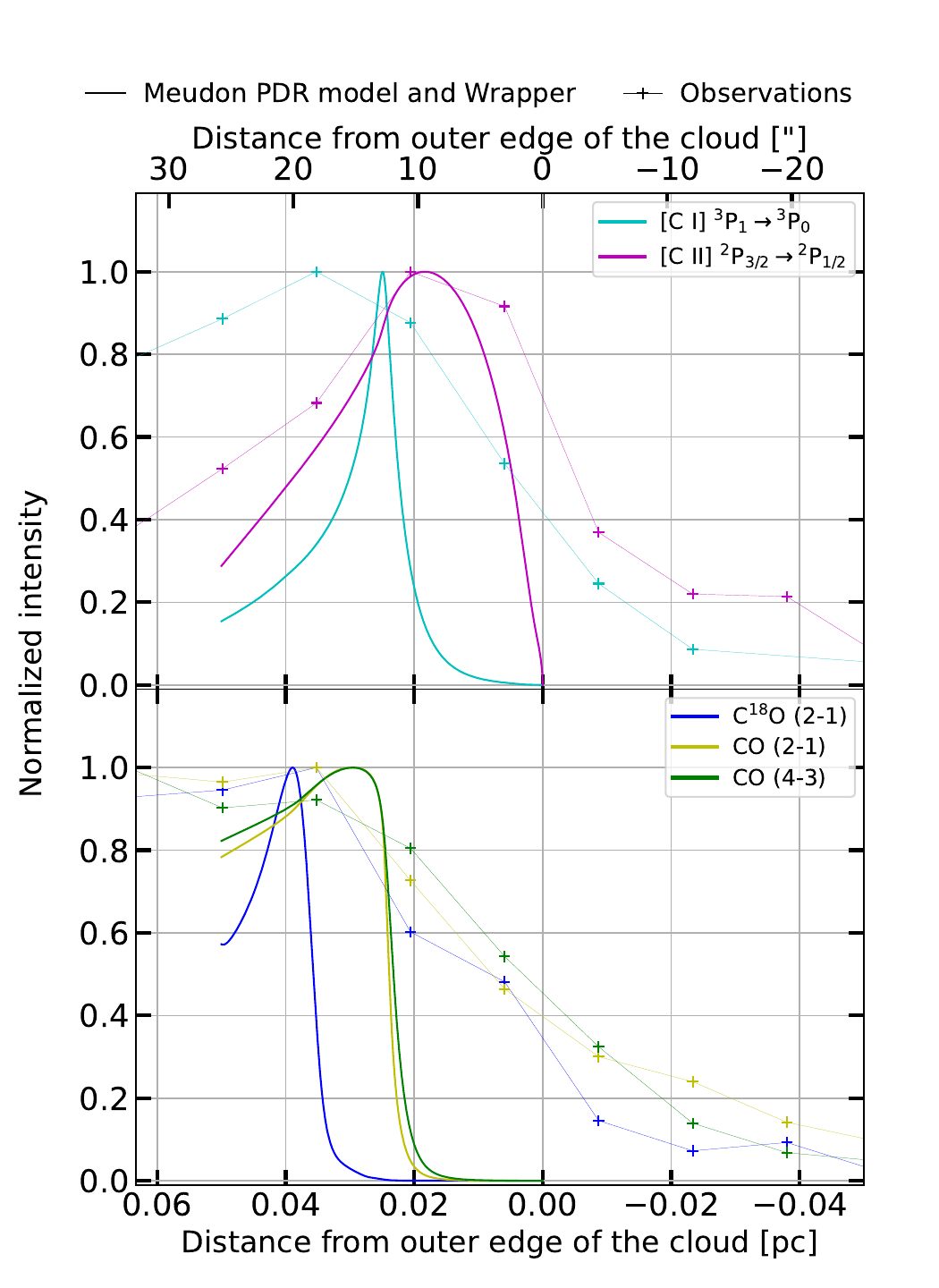}
\includegraphics[width=0.45\textwidth]{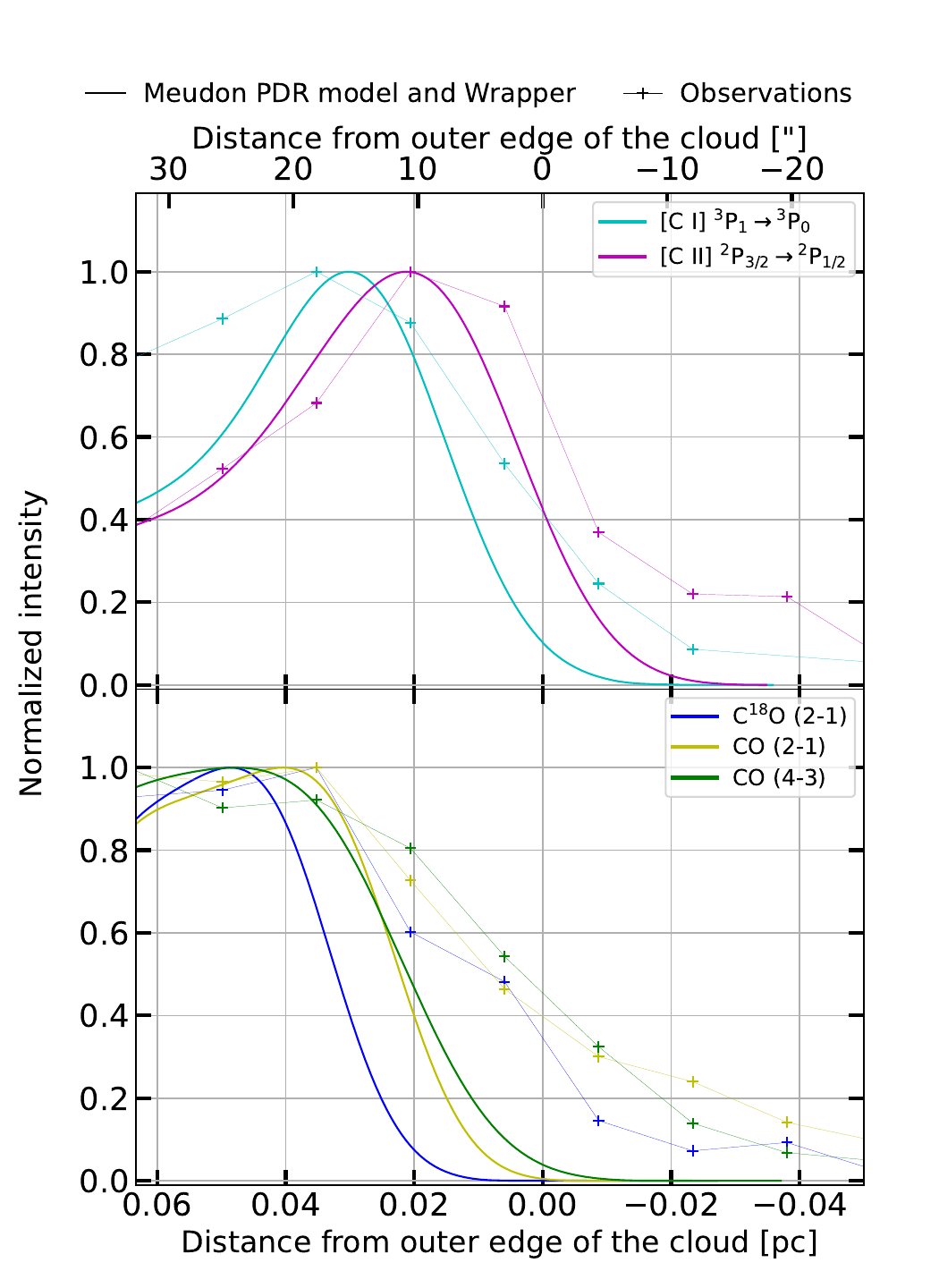} 
\includegraphics[width=0.93\textwidth]{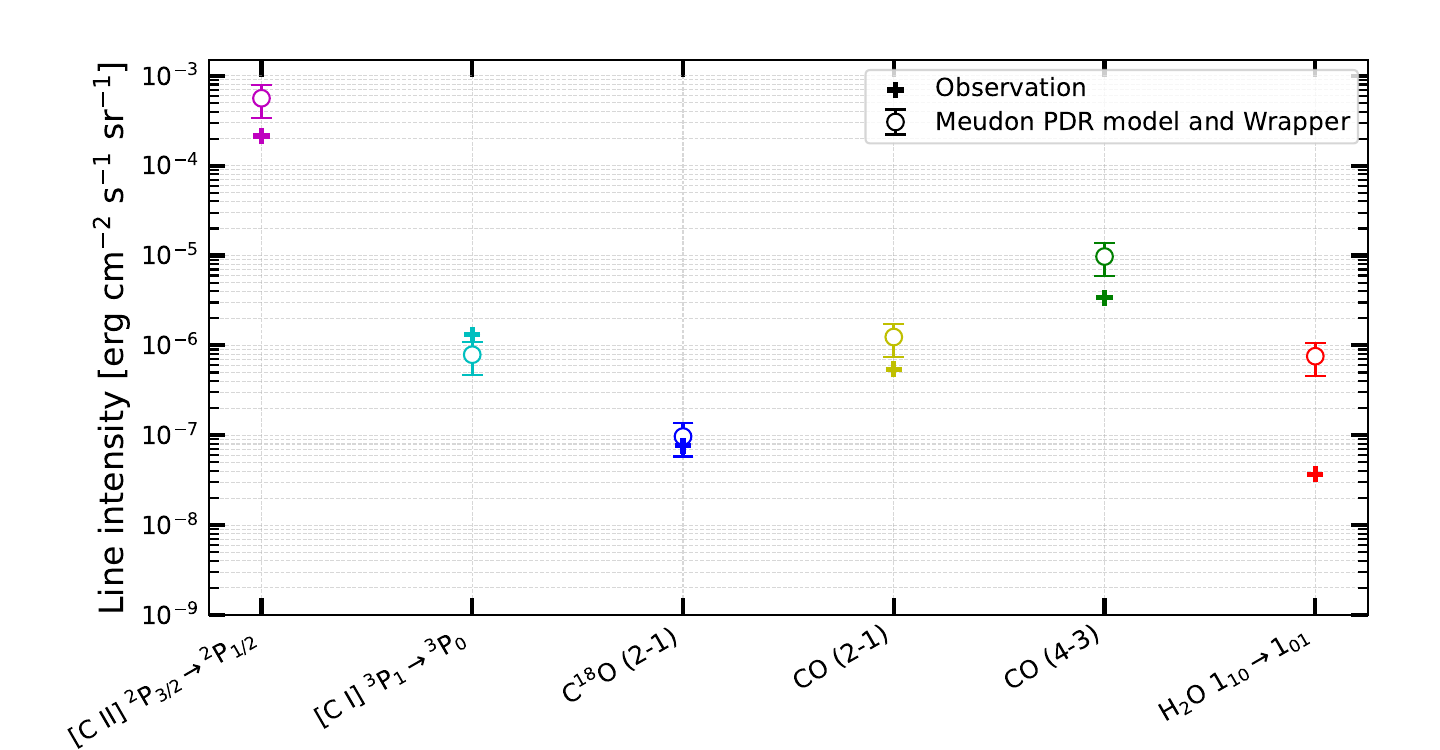} 
\caption{(Top) Spatial profiles of normalized emission line
  intensities observed in the Horsehead nebula, compared with
  predictions from a Meudon PDR model with
  $P_{\mathrm{th}} = 4 \times 10^6$~K\,cm$^{-3}$ and $G_0 = 100$,
  wrapped in a locally spherical cloud with a curvature radius
  $R_{\mathrm{C}} = 0.05$~pc. These conditions correspond to commonly
  adopted literature values. The left panel shows the raw profiles
  from the wrapping procedure, while the right panel presents the same
  profiles convolved to the native angular resolution of each tracer.
  On those figures, the radiation field comes from the right. (Bottom)
  Observed maximum intensities of various tracers (colored crosses)
  compared to model predictions (circles with error bars). All tracers
  are well reproduced, except for H$_2$O, where the model
  overestimates the observed maximum intensity by an order of
  magnitude.}
\label{fig:best_model}
\end{figure*}

\begin{figure*}
\centering
\includegraphics[width=0.9\textwidth]{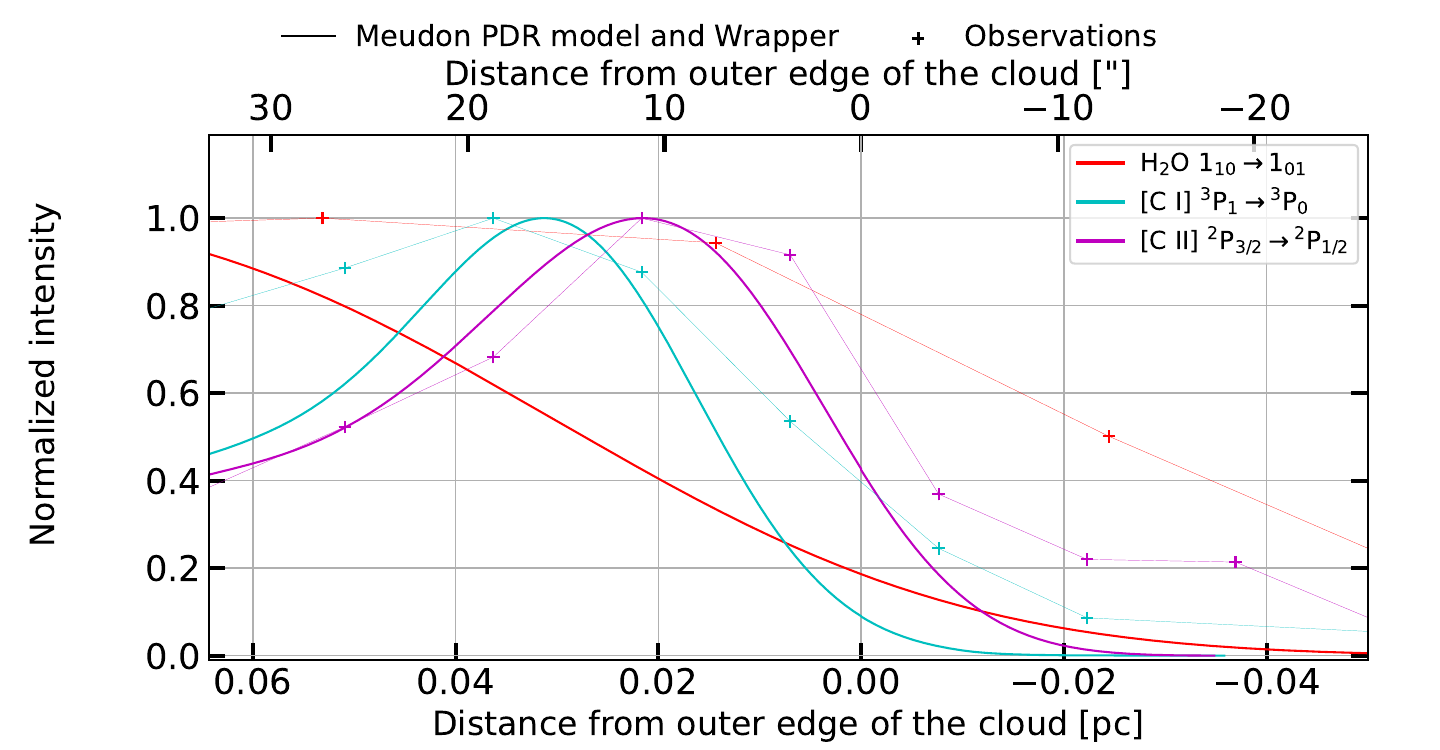}
\caption{Spatial profiles of normalized emission line intensities
  observed in the Horsehead nebula, compared to predictions from
  Meudon PDR model with $P_{\mathrm{th}} = 4 \times 10^6$~K\,cm$^{-3}$
  and $G_0 = 100$, wrapped in a locally spherical cloud with curvature
  radius $R_{\mathrm{C}} = 0.05$~pc. The profiles are convolved to the
  native angular resolution of the tracer maps. On this figure, the
  radiation field comes from the right. We see a significant shift
  between the observed H$_2$O profile and the model, the latter being
  much deeper in the cloud. }
\label{fig:best_model_H2O}
\end{figure*}

Figure~\ref{fig:best_model} and \ref{fig:best_model_H2O} show the
spatial emission profiles (after the wrapping procedure, and
convolution to the native angular resolution of each tracer) and the
maximum intensities including the H$_2$O $1_{10}$--$1_{01}$ line, for
the model : $P_{\mathrm{th}} = 4 \times 10^6$~K\,cm$^{-3}$,
$G_0 = 100$, and $R_{\mathrm{C}} = 0.05$~pc. For water, the model
predicts an intensity about 20 times higher than the
observed value, and a spatial distribution of the emission much more
recessed into the cloud. Models with lower $G_0$ values provide
a somewhat better fit to the observations. However, the model water
line intensity is stil a factor of 7 higher than the observations (see
Fig.~\ref{fig:app_literature_model} in
Appendix~\ref{app:Appendix_Indep_to_G0}). 

There are several possible explanations for these discrepancies.
First, H$_2$O excitation is very sensitive to radiative transfer
effects \citep{Poelman2006, Gonzalez2008}, such as non-local pumping
by warm dust emission. Also, the line is strongly optically thick, 
with optical depths of the order of $10^{3}$,
so the emergent intensity is determined by the local excitation
temperature at the $\tau \sim 1$ surface. The location of this surface
is highly sensitive to both the H$_2$O abundance profile and the line
width, which is controlled by turbulent broadening. A smaller
turbulent velocity leads to a narrower line, increasing the opacity,
and thus shifting the $\tau \sim 1$ surface outward, closer to the
cloud edge where the excitation temperature is lower, thereby
decreasing the emergent intensity.

To test the sensitivity of the results to turbulent velocity
dispersion, we also considered the smaller value of
$\sigma_{\mathrm{turb}} = 0.225$~km\,s$^{-1}$ derived from the
analysis of \citet{Gerin2009}, who derived this value for the
Horsehead from HCO observations combining Plateau de Bure
Interferometer (PdBI) and IRAM-30m data. In this case the predicted
intensity decreases by about 24\%, from
$7.58 \times 10^{-7}$~erg~cm$^{-2}$~s$^{-1}$~sr$^{-1}$ to
$5.75 \times 10^{-7}$~erg~cm$^{-2}$~s$^{-1}$~sr$^{-1}$. Both
predictions remain far above the observed value of
$3.67 \times 10^{-8}$~erg~cm$^{-2}$~s$^{-1}$~sr$^{-1}$. For the other
tracers, we find that decreasing the turbulent velocity has almost no
effect on [C\,\textsc{ii}] and [C\,\textsc{i}], but it reduces the CO
line intensities by $\sim 20\%$, thereby slightly improving the
agreement with the observations. 

In addition, the intensity of optically thick lines varies non
linearly with the abundance profile. A small change in the predictions
of the density of H$_2$O can result in significant variations in line
intensities. We checked that our model is perfectly in line with the
classical PDR model of the chemistry of H$_2$O \citep{Hollenbach2009},
except that surface chemical data (e.g., binding energies) have been
updated and the modelling of some processes has been modernized
\citep{Cuppen2017}. Also, we know that the model reproduces H$_2$O
line intensities in more intense PDRs such as the Orion Bar
\citep{Putaud2019}. This may suggest that particular processes may be at
work in the Horsehead, possibly involving grains with unusual
properties, whether in composition or size distribution.

The detailed analysis of dust and H$_2$ emission recently observed by
JWST in the Horsehead nebula could help clarify this issue
\citep{Abergel2024}. These findings also highlight the need to refine
the modelling of chemical processes on interstellar grains,
particularly by acknowledging that the physical conditions of these
grains likely differ from those under which binding energies and
process efficiencies, such as photo-desorption yields, are measured in
laboratory experiments \citep{Hacquard2024}. For instance, given
the diversity of interstellar grain types, it would be more realistic
to adopt distributions of binding energies for adsorbed species
instead of single fixed values. However, this would require the
development of new formalisms to accurately simulate surface
chemistry. Last but not least, another factor to consider is the
geometry of the emission and non-local radiative transfer effects: the
discrepancy between models and observations may well arise from
scattering of the radiation emitted by water molecules, as discussed
in the following section.

\section{Scattered $o$-H$_2$O 1$_{10}$-1$_{01}$ 557\,GHz line emission 
  by low-density warm envelopes}\label{sec:scattering}

In this section, we describe a plausible scenario explaining why the
$o$-H$_2$O 1$_{10}$-1$_{01}$ 557\,GHz line emission is more extended
toward the rim of the PDR than the $^{12}$CO (2--1) emission, and why
it nearly coincides with the [CII]\,158\,$\mu$m emission when
convolved to the same low angular resolution ($\sim$38$''$), which
traces lower-density gas at the rim of the PDR
\citep{Pabst2017,Bally2018}. Firstly, we note that a steep gas density
gradient exists in the Horsehead, with densities varying from the
PDR/H\,\textsc{ii} interface ($n_{\rm H} \approx$ a few
$10^3$\,cm$^{-3}$, $T \gtrsim 100$\,K) to the more shielded, cold
molecular cores ($n_{\rm H} \approx$ a few $10^5$\,cm$^{-3}$,
$T \approx 20$\,K) \citep[e.g.,][]{Habart2005,Pety2007,Gerin2009,
  Goicoechea2009, Hernandez2023}. The nearly edge-on orientation of
the PDR allows these different environments to be partially spatially
resolved.

\begin{figure*}
\centering
\includegraphics[width=0.82\textwidth]{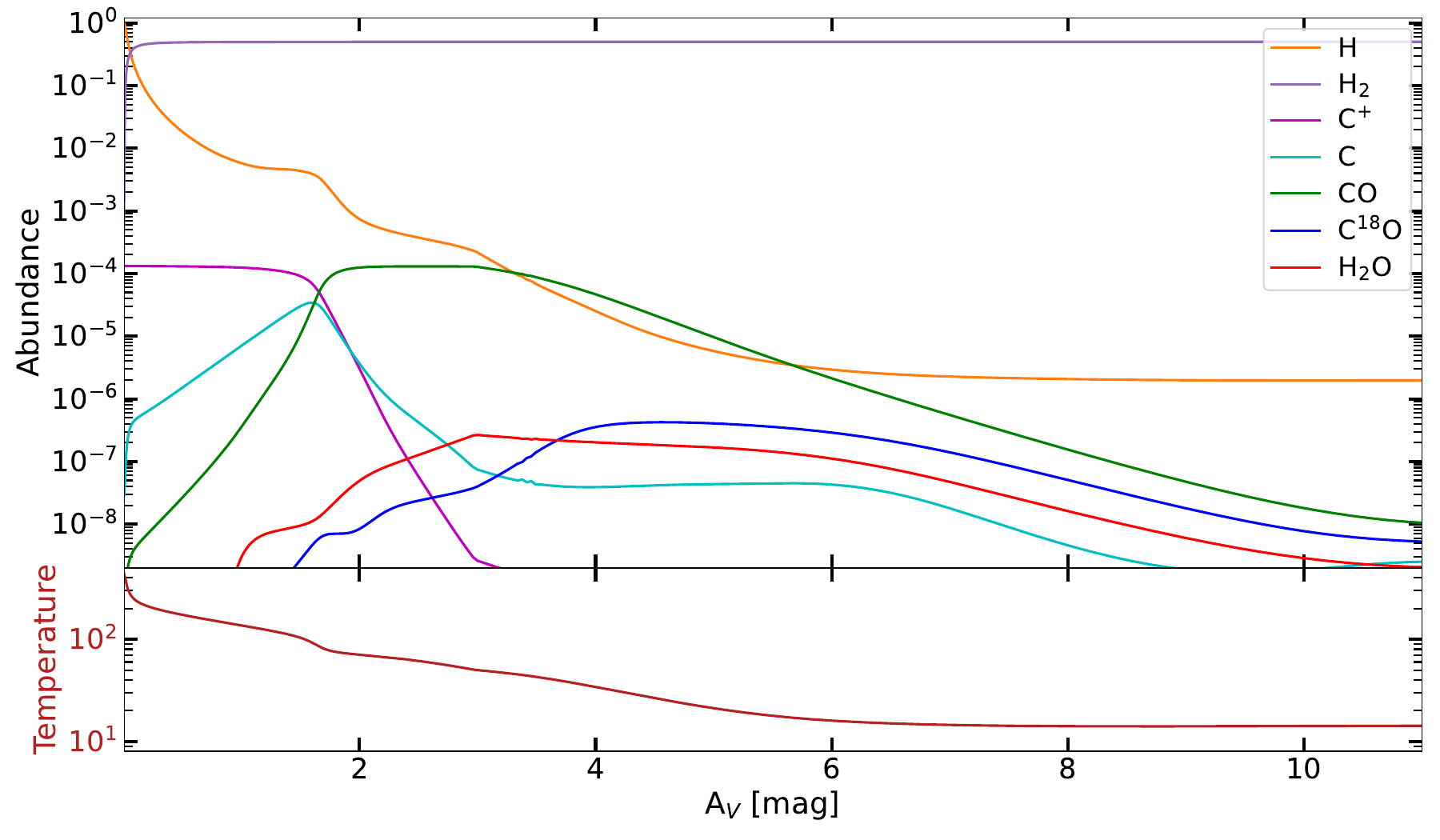} 
\caption{ Abundances of key species with respect to hydrogen
  nuclei (top panel) and gas temperature (bottom panel) profiles of
  the Horsehead nebula model with
  $P_{\mathrm{th}} = 4 \times 10^6$~K\,cm$^{-3}$, $G_0 = 100$,
  computed with the Meudon PDR code (no wrapping or convolution
  procedure is applied to these profiles). We see the H/H$_2$
  transition taking place at around $A_V$\,$\sim$\,0.06\,mag, the
  C$^+$/C/CO transition at $A_V$\,$\sim$\,1.7\,mag, and the peak of
  H$_2$O abundance is reached at an $A_V$ value of 3\,mag.}
\label{fig:edge-on}
\end{figure*}

Our PDR models (Sect.~\ref{sec:H2O_Wrapper}) predict that the peak of
the water vapor abundance, reaching $2.6 \times 10^{-7}$, occurs at
$A_V$\,$\sim$\,3\,mag, well beyond the CO/C transition layer of the
PDR, as can be seen on Figure \ref{fig:edge-on} showing the edge-on
Meudon PDR model output (thus without the wrapping). As a reference,
models from \cite{Putaud2019} yielded a comparable abundance of H$_2$O
of $2 \times 10^{-7}$, that occurred at $A_V$\,$\sim$\,9\,mag in the
Orion Bar, due to the substantially stronger radiation field. Our water
abundance peak roughly corresponds to the point where the rate of
gas-phase O atoms sticking to grains equals the photo-desorption rate
of water molecules from icy grain surfaces. At this depth, the gas
temperature is approximately 40\,K, and the H$_2$O peak lies deeper
than the $^{12}$CO abundance peak. In this region of the PDR, the
H$_2$O abundance profile closely resembles that of C$^{18}$O. Hence,
contrary to observations, one would expect the $o$-H$_2$O 557\,GHz
emission to be as extended as the C$^{18}$O\,(2--1) emission. In
contrast to C$^{18}$O, however, H$_2$O starts to form close to the PDR
surface, ahead of the CO/C transition, where the C$^+$ abundance is
much greater than that of CO, temperatures are high, and densities are
lower than at the H$_2$O abundance peak. With this in mind, one can
approximate the H$_2$O line radiative transfer problem in the
Horsehead using a two-component model: a denser, moderately cold PDR
component with a large H$_2$O column density (in the line of sight),
surrounded by a lower-density, warm envelope with a lower H$_2$O
column density. This envelope represents the bright [CII]158~$\mu$m
but faint $^{12}$CO\,(2--1) emission edge of the PDR
\citep[e.g.,][]{Pabst2017,Hernandez2023}.

Owing to the very high critical density for collisional excitation,
\mbox{$n_{\rm cr}$(1$_{10}$--1$_{01}$)\,$\simeq\,10^8$\,cm$^{-3}$},
and the elevated 557\,GHz line opacity for moderate H$_2$O column
densities, the excitation of the 1$_{10}$--1$_{01}$ transition is very
subthermal ($n_{\rm H} \ll n_{\rm cr}$) (see, e.g.,
\citealt{Flagey2013}), which implies \mbox{$T_{\rm ex} \ll T$}, and is
prone to non-local radiative effects such as line resonant scattering
by low-density envelopes containing water vapor. In such envelopes,
the local $o$-H$_2$O\,557\,GHz line emission is negligible because
\mbox{$T_{\rm ex}$\,$\rightarrow$\,$T_{\rm cmb}$}. However, when
$o$-H$_2$O 557\,GHz line photons emitted by the deeper PDR layers
diffuse through the envelope, they are self-absorbed and
\mbox{re-emitted} by water molecules, which increases $T_{\rm ex}$ in
the \mbox{low-density gas} and scatters the emission to greater
distances. This produces an impression of a more extended spatial
distribution of H$_2$O emission. While this scattering process has
been invoked to explain the extended distribution and anomalous
\mbox{$J$\,=\,1--0} line intensity ratios of very polar molecules such
as HCO$^+$ or HCN in {dark clouds \citep[i.e., rotational transitions with a high
$n_{\rm cr}$;
][]{Langer1978,Walmsley1982,Cernicharo1984,Gonzalez1993,Goicoechea2022},
all these studies required cold envelopes. The interesting property of
H$_2$O is that the critical density of the ground-state
1$_{10}$-1$_{01}$ transition is at least 3 orders of magnitude higher
than that of HCN and HCO$^+$ \mbox{$J$\,=\,1--0}. As we demonstrate
below, this implies that the excitation temperature of the
1$_{10}$--1$_{01}$ transition remains very low even at high gas
temperatures, provided that the gas density is significantly lower
than the effective critical density of this transition. In this case,
a warm/hot envelope containing moderately water vapor abundances will
produce self-absorption and resonant scattering due to very low
excitation temperature of the $1_{10}-1_{01}$ in the envelope. This
non-local radiative effect cannot be reproduced by standard (local)
escape probability methods.

To investigate the impact of line scattering on the emergent H$_2$O
line intensities, we use a \mbox{non-local} and non-LTE Monte Carlo
radiative transfer code \citep[see][]{Goicoechea2006,Goicoechea2022}
and develop a “simple” two-component model of the water vapor line
emission adapted to the physical conditions known to exist in the
Horsehead PDR. We used the H$_2$O--H$_2$ inelastic collision rate
coefficients of \cite{Daniel2011,Daniel2012}. Following a large list
of observational studies of this source, we adopt
\mbox{$n_{\rm H}$=6$\times$10$^4$\,cm$^{-3}$} and \mbox{$T$\,=\,40\,K}
for the dense PDR component, and
\mbox{$n_{\rm H}$=6$\times$10$^{3}$\,cm$^{-3}$}, $T$\,=\,60\,K, and
for the extended envelope near the PDR/HII~region interface. We found
a good match to the observed line intensities by adopting a
(beam-averaged) line-of-sight $o$-H$_2$O column densities,
\mbox{$N$($o$-H$_2$O)\,$\simeq$\,4.3$\times$10$^{13}$\,cm$^{-2}$} and
\mbox{$N$($o$-H$_2$O)\,$\simeq$\,1.7$\times$10$^{13}$\,cm$^{-2}$} in
these two components, respectively. These models include thermal,
microturbulent, and line opacity broadening. To reproduce the observed
line-widths, we adopt $\sigma_{\rm turb}$\,$\simeq$\,0.4\,km\,s$^{-1}$
in the dense PDR and $\sigma_{\rm turb}$\,$\simeq$\,1\,km\,s$^{-1}$ in
the extended envelope. An increased velocity dispersion in the
envelope diminishes the depth of the 557\,GHz self-absorption line
dip. Figure~\ref{fig:mtc_models_1} shows the resulting line and radial
\mbox{$T_{\rm ex}$\,(1$_{10}$-1$_{01}$)} profiles. The blue curves
represent a model of the dense PDR component-only, while the red
curves represent a model of the low-density envelope-only. The black
curves represent the two-component model.

\begin{figure}
   \centering
   \includegraphics[scale=0.5, angle=0]{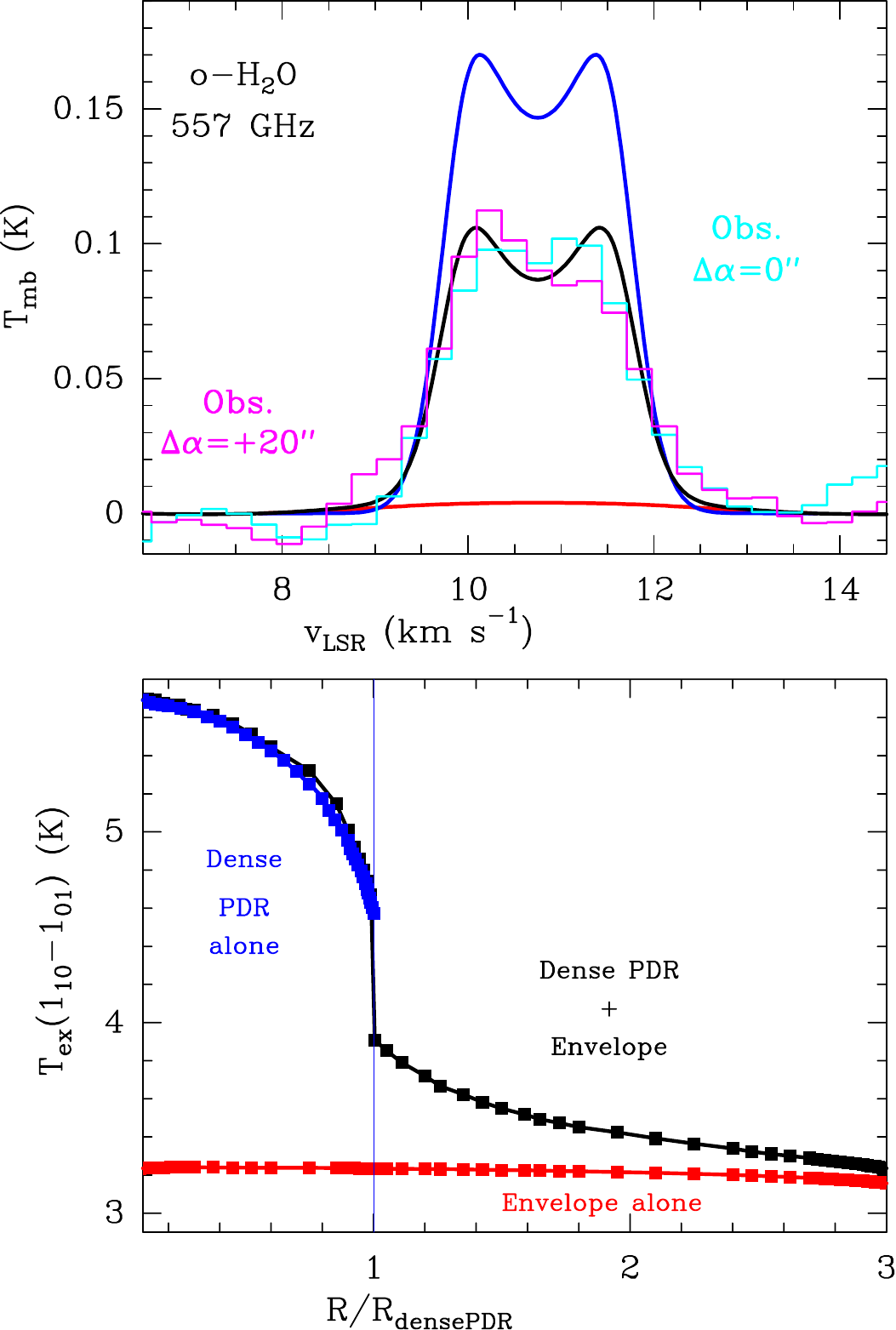}   
   \caption{Two-component, non-local radiative transfer models adapted
     to the physical conditions in the Horsehead. The upper panel
     shows the \mbox{$o$-H$_2$O\,557\,GHz} synthetic line profiles
     (continuous curves) and observed line profiles (histograms) at
     the emission peak. The lower panel shows the predicted excitation
     temperatures of the \mbox{1$_{10}$-1$_{01}$} transition. The blue
     curves represent the model of the dense PDR component alone. The
     red curves represent the model of the low-density envelope alone,
     while the black curves represent the two-component model. }
  \label{fig:mtc_models_1}
\end{figure}

Due to the low gas density, the envelope-only model shows highly
subthermal (\mbox{$T_{\rm ex}$\,$\simeq$\,3.2\,K}) line emission,
despite the moderate gas temperatures. Thus, the predicted $o$-H$_2$O
557\,GHz line intensities are very faint (a few mK) and would remain
undetectable by Herschel/HIFI. The dense-PDR-only alone, with and
order of magnitude higher gas density and a lower temperature (closer
to the energy of the upper level 1$_{10}$), leads to more efficient
collisional excitation and thus increased excitation temperatures
\mbox{($T_{\rm ex}$\,$\sim$4.5--5.5\,K)}. In addition, the $o$-H$_2$O
557\,GHz line from the dense component is optically thick, which
produces line opacity broadening and line-trapping (i.e., an increase
of $T_{\rm ex}$ at low radii despite the gas density is constant).
Compared to the envelope-only model, the two-component model (black
curves) has different excitation conditions in the envelope region
\mbox{($R > R_{\rm densePDR}$)}. The absorption and reemission of
$o$-H$_2$O\,557\,GHz line photons has two consequences: ($i$) It
reduces the apparent $o$-H$_2$O\,557\,GHz line intensity toward the
dense component, and ($ii$) it increases
\mbox{$T_{\rm ex}$(1$_{10}$-1$_{01}$)} in the envelope to levels that
produce detectable scattered $o$-H$_2$O\,557\,GHz line emission at
$R > R_{\rm densePDR}$. This is a purely radiative effect and leads to
$o$-H$_2$O\,557\,GHz line emission beyond the dense PDR component.
This scenario would imply that our detection of the 
\mbox{$o$-H$_2$O\,557\,GHz} emission at the rim of the PDR is
scattered emission from deeper layers of the PDR.

\begin{figure}[h!] 
  \centering 
  \includegraphics[scale=0.49, angle=0]{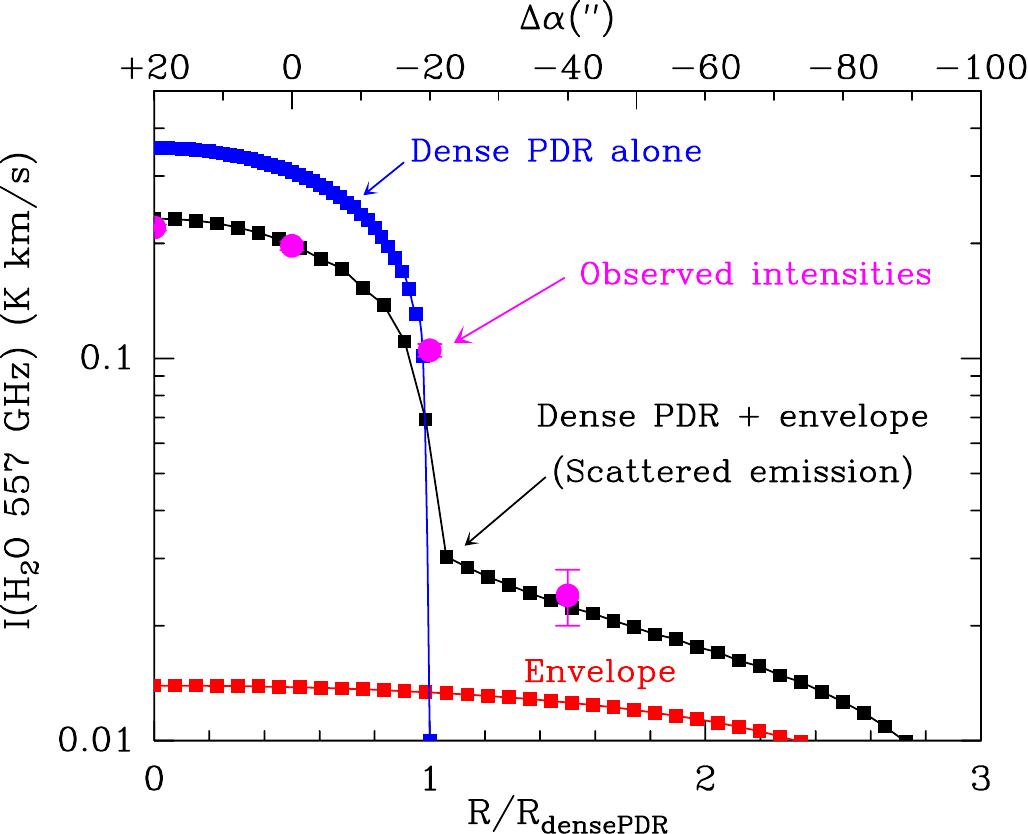}  
  \caption{Same as the lower panel of Fig.~\ref{fig:mtc_models_1} but
    showing the predicted radial \mbox{$o$-H$_2$O 1$_{10}$-1$_{01}$}
    integrated line intensities. The magenta-filled circles represent
    the observed line intensities across the Horsehead PDR, with the
    peak position at $\Delta \alpha = +20''$.
  } \label{fig:mtc_models_2}
\end{figure}

Figure~\ref{fig:mtc_models_2} shows the predicted radial $o$-H$_2$O
1$_{10}$-1$_{01}$ integrated line intensities, with the black curves
representing the two-component model. This model results in detectable
\mbox{$o$-H$_2$O\,557\,GHz} emission that is more extended than the
dense PDR region. It also results in an decrease in the
\mbox{$o$-H$_2$O\,557\,GHz} line intensity for line of sights toward
the core (e.g., for $R/R_{\rm densePDR}= 0$). This simple
    two-component model reasonably fits the observed
    \mbox{$o$-H$_2$O\,557\,GHz} line intensities across the PDR
    (magenta-filled circles in \mbox{Fig.~\ref{fig:mtc_models_2}}).
We recall that a necessary condition for this radiative effect to
occur is the presence of an extended, low-density envelope
(lower excitation) containing moderate amounts of water
vapor, as predicted by our PDR photochemical model. The size of the
H$_2$O scattered emission region and the degree of line
self-absorption ultimately depends in the exact physical conditions
and water vapor abundance in the different components. We note
    that resonant scattering does not affect the excitation or
    emission of low–critical-density transitions (e.g.,~low-$J$ CO,
    \CII, or \CI~lines). Indeed, most of the C$^{18}$O~2--1 emission
    originates in the dense PDR interior, while any C$^{18}$O present
    in the envelope will be excited collisionally
    \mbox{($T_{\rm ex}$\,$\simeq$\,$T_{\rm k}$)} and not radiatively.
    As the line emission from the dense PDR is optically thin, this
    emission is not self-absorbed. For C$^+$, the \CII\,158\,$\mu$m
    line (also collisionally excited) arises mainly from the
    foreground envelope. These two cases, both with $n_{\rm cr}$ of a
    few 10$^3$\,cm$^{-3}$, are presented in
    Appendix~\ref{app:Appendix_CO_Cplus_toy}.

In summary, our simple model captures the average physical
    conditions and H$_2$O column densities within the large
    Herschel/HIFI beam. Higher angular resolution observations of
    water vapor will be necessary to properly resolve the size of each
    component and to account for the gradients in physical conditions
    and water abundance.

\section{Summary and conclusions}
\label{sec:summary}

We have analyzed archival \emph{Herschel} observations of water
emission toward the Horsehead PDR, along with supporting ground-based
and airborne observations of CO isotopologues and fine structure lines
of ionized and atomic carbon. The main results of this study can be
summarized as follows:

\begin{itemize}

\item Water emission in the Horsehead nebula is very weak and,
  surprisingly, extends outward beyond other PDR tracers, such
  as CO isotopologues or [C\,{\sc i}]~609~$\mu$m, and as far out as
  [C\,{\sc ii}]~158\,$\mu$m. 

\item We developed a PDR wrapper to model line emission from 3D
  objects illuminated by a flux of UV photons. PDR modeling
  shows that the [C\,{\sc ii}], [C\,{\sc i}], and CO isotopologue
  intensities and spatial profiles provide strong constraints on the
  thermal pressure ($P_{\mathrm{th}} = 4 \times 10^6$~K\,cm$^{-3}$),
  but not on $G_0$. A model with published source parameters
  provides a reasonable fit to the data, while underestimating the
  intensities of the [C\,{\sc ii}] and CO $J=4-3$ lines by a factor of
  $\sim 2.5$.
  Spatial profiles are also well reproduced, except for
  CO isotopologues, where the increase on the illuminated side of the
  PDR is steeper than observed.
  
\item Water vapor abundance in the PDR model reaches
  $3.6 \times 10^{-7}$ at $A_V \sim 3$ mag. However, the ground state
  $o$-H$_2$O 557 GHz line is systematically overestimated by the
  models by at least a factor of 7 for any values of the model
  parameters. This line has a very high optical depth,
  $\sim 10$, and the emergent line intensity is sensitive to radiative
  transfer effects and the assumed turbulent line width. Nevertheless,
  the PDR modeling suggests that some ingredients describing the
  processes that lead to the formation and destruction of water ice on
  grain surfaces are missing in the model.

\item We demonstrate that scattering of photons by water molecules in
  a low-density envelope surrounding the dense PDR provides a
  plausible explanation for the observed spatial extent of the water
  emission. While an isobaric Meudon PDR model naturally
  includes a low-density, warm envelope, the local escape and
  radiative transfer formalism in the code does not treat correctly
  the resonant scattering of water photons by the outer, lower-density
  gas layers. 
  
\end{itemize}

\begin{acknowledgements}
  
  This research was carried out at the Jet Propulsion Laboratory,
  California Institute of Technology, under a contract with the
  National Aeronautics and Space Administration (80NM0018D0004).
  D.C.L. acknowledge financial support from the National Aeronautics
  and Space Administration (NASA) Astrophysics Data Analysis Program
  (ADAP). HIFI was designed and built by a consortium of institutes
  and university departments from across Europe, Canada, and the
  United States (NASA) under the leadership of SRON, Netherlands
  Institute for Space Research, Groningen, The Netherlands, and with
  major contributions from Germany, France, and the US. This research
  has made use of the NASA/IPAC Infrared Science Archive, which is
  funded by the National Aeronautics and Space Administration and
  operated by the California Institute of Technology. JRG thanks the
  Spanish MCINN for funding support under grant PID2023-146667NB-I00.
  This work was supported by the Thematic Action “Physique et Chimie
  du Milieu Interstellaire” (PCMI) of INSU Programme National “Astro”,
  with contributions from CNRS Physique \& CNRS Chimie, CEA, and CNES.
  This research was achieved using the POLLUX database
  (pollux.oreme.org) operated at LUPM (Université de Montpellier -
  CNRS, France) with the support of the PNPS and INSU. V.M. and J.R.G.
  thank the Spanish MCINN for funding support under grant
  PID2023-146667NB-I00. We thank an anonymous referee for helpful
  comments.
\end{acknowledgements}

\bibliographystyle{aa} 
\bibliography{biblio}  

\appendix       

\section{\label{Appendix_PDR_Wrapper}Presentation of the PDR wrapper} 
\label{app:wrapper}

\subsection{Aims of the code}
    
The Meudon PDR code \citep{LePetit2006} models interstellar clouds as
1D, plane-parallel slabs, infinite and homogeneous along their
surface. It computes the steady-state gas and chemistry structure
along the normal to the surface, as a function of optical depth, and
predicts line intensities emerging from the illuminated side. Level
excitation is determined at each position in the cloud by solving the
non-LTE problem, taking into account radiative and collisional
transitions, as well as chemical formation and destruction in specific
levels. The algorithm that computes the stationary level populations
is intimately coupled to the solution of the radiative transfer
equation, as described in \citet{Gonzalez2008}. The algorithm
considers non local pumping in the continuum (dust emission plus CMB).
For example, the radiation field emitted by warm dust at the edge of
the PDR can pump transitions deeper in the cloud where dust is cold.
Concerning line emission, the algorithm uses a \textit{on the spot}
approximation meaning that the line can be re-absorbed, but only
locally. For H$_2$O, the code considers inelastic collisions with H
\citep{Daniel2015}, He (scaling of the collision rates with H), ortho
and para H$_2$ \citep{Daniel2011}, e$^-$ \citep{Faure2004}. 

Edge-on clouds illuminated from the side reveal PDR stratification
through spatially offset line emissions. Although their complex
geometry requires simplifications, the Meudon code allows for a
detailed treatment of microphysics, chemistry, and radiative transfer.
Extending this approach to a full 2D model with the same level of
refinement in the treatment of physical processes would, however, be
computationally prohibitive.

When the PDR width is small compared to the cloud radius, radiative
transfer can be simplified by mapping the 1D model onto a 2D spherical
geometry. This is the basis of the PDR Wrapper, which reconstructs
line emission by solving radiative transfer along each line of sight
through a spherical cloud cross-section.

\subsection{Geometry}

We assume a locally spherical curvature for the cloud surface, with
center of curvature $O$ and radius $R_{\text{C}}$. The PDR is modeled
as the outermost shell of thickness $d_{\text{PDR}} \ll R_{\text{C}}$.

The radiation field is assumed constant and perpendicular to the line
of sight. The cloud has a curvature radius $R_{\text{C}}$ and a
constant horizontal thickness $d_{\text{PDR}}$. It is modeled as the
shell between two arcs of radius $R_{\text{C}}$, with centers shifted
horizontally by $d_{\text{PDR}}$ (Fig.~\ref{fig:Horizontal_geometry}).
For a point on the outer arc making an angle $\alpha$ with the
horizontal, the depth $d$ from the illuminated edge is:

\begin{equation}
  d = R_{\text{C}}(\cos{\alpha} - 1) + b,
\end{equation}
\noindent with $\sin{\alpha} = s/R_{\text{C}}$. The expression for
$s_{\text{max}}$ remains unchanged, while: 
\begin{equation}
            s_{\text{min}}(b) = 
            \begin{cases}
                0 & \text{if } b < d_{\text{PDR}}, \\
                \sqrt{R_{\text{C}}^2 - (R_{\text{C}} - b + d_{\text{PDR}})^2} & \text{otherwise}.
            \end{cases}
\end{equation}

\begin{figure}
\centering
\includegraphics[width=0.99\linewidth]{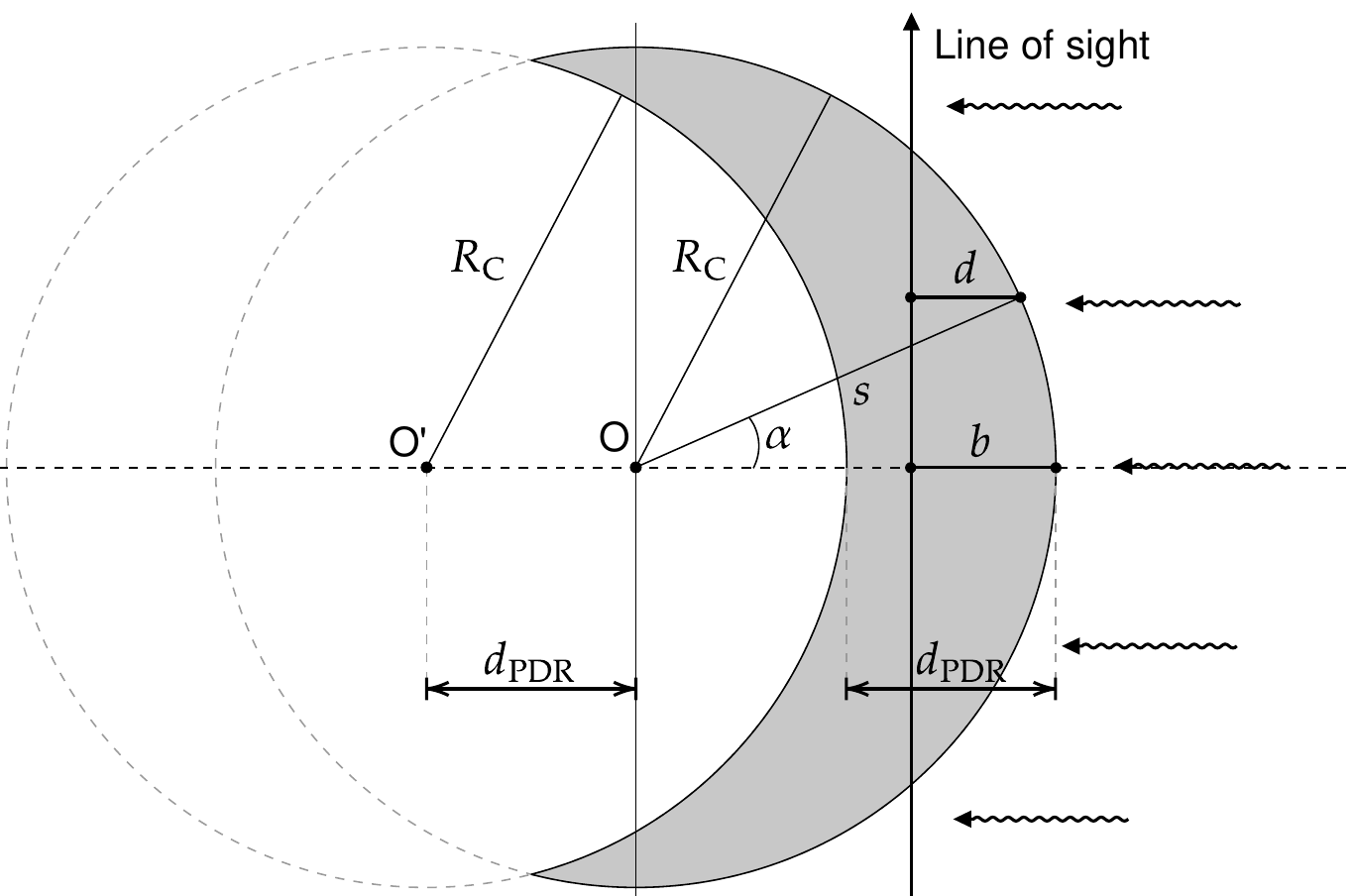}
\caption{Horizontal geometry used for the wrapping of Meudon PDR
  models. The beamed ISRF remains perpendicular to the line of sight.}
\label{fig:Horizontal_geometry}
\end{figure}

\subsection{Radiative transfer}

Each image pixel integrates emission and absorption along a single
line of sight. Our model solves the radiative transfer equation along
multiple, parallel, and independent lines of sight across the cloud.

The goal is to compute the intensity of a user-selected line at
frequency $\nu_{ul}$, where $u$ and $l$ are the upper and lower energy
levels, respectively. Accounting for spontaneous emission ($A_{ul}$),
stimulated emission ($B_{ul}$), and absorption ($B_{lu}$), the
radiative transfer equation reads:
\begin{equation}
            \frac{\mathrm{d}I_{\nu}}{\mathrm{d}s} = \frac{\phi(\nu,
              \nu_{ul}, s)}{4 \pi} h \nu \left[A_{ul} n_u(s) +
              \left(B_{ul} n_u(s) - B_{lu} n_l(s)\right) I_{\nu}
            \right], 
\end{equation}
\noindent where $n_u(s)$ and $n_l(s)$ are the level populations along
the line of sight. 

The normalized line profile $\phi$ is assumed Doppler-broadened due to
thermal motions and turbulence, and is given by:
\begin{align}
            \phi(\nu, \nu_{ul}, s) &=
                                     \frac{1}{\sigma_{\nu_{ul}}(s)\sqrt{2\pi}}
                                     \exp\left[-\frac{(\nu -
                                     \nu_{ul})^2}{2
                                     \sigma^2_{\nu_{ul}}(s)}\right],
  \\ 
            \sigma_{\nu_{ul}} &= \frac{\sqrt{2}}{2} \frac{\nu_{ul}}{c}
                                \sqrt{\frac{2k_{\text{B}}T}{m} +
                                \sigma_{\text{turb}}^2}, 
\end{align}
\noindent where the normalization relation is:
\begin{equation}
            \int_0^{+\infty} \phi(\nu, \nu_{ul}, s)\,\mathrm{d}\nu = 1.
\end{equation}

The Einstein relations (valid when working with $I_\nu$) are:
\begin{equation}
            B_{ul} = \frac{c^2}{2h\nu_{ul}^3} A_{ul}, \quad B_{lu} =
            \frac{g_u}{g_l} B_{ul}. 
\end{equation}

Rewriting the transfer equation as:
\begin{equation}
            \frac{\mathrm{d}I_{\nu}}{\mathrm{d}s} = \epsilon_{\nu} -
            \alpha_{\nu} I_{\nu}, 
\end{equation}
\noindent where we define the emissivity and absorption coefficients:
\begin{align}
            \epsilon_{\nu} &= \frac{\phi(\nu, \nu_{ul}, s)}{4\pi} h
                             \nu A_{ul} n_u(s), \\ 
            \alpha_{\nu} &= \frac{\phi(\nu, \nu_{ul}, s)}{4\pi} h \nu
                           \left[B_{lu} n_l(s) - B_{ul} n_u(s)\right], 
\end{align}

The formal solution is:
\begin{equation}
            I_{\nu}(s) = \int_{s_{\text{min}}}^{s}
            \epsilon_{\nu}(s')\, e^{-\tau_{\nu}(s')} \,\mathrm{d}s',  \quad 
            \tau_{\nu}(s) = \int_{s_{\text{min}}}^{s}
            \alpha_{\nu}(s')\, \mathrm{d}s'. 
\end{equation}

Finally, the total emergent intensity along the line of sight (impact
parameter $b$) is obtained by frequency integration: 
\begin{equation} 
            I(b) = \int_0^{+\infty} I_{\nu}(b) \, \mathrm{d}\nu.
\end{equation}

\subsection{Convolution with instrumental PSF}

The PSF (point spread function) characterizes the response of an
instrument to a point-like source. It defines the instrument angular
resolution. To model the instrumental response to the spatial
profiles, we convolve $I(b)$ or $N(b)$ with a Gaussian kernel:

\begin{equation}
            g(x) = \frac{1}{\sigma \sqrt{2 \pi}} \, \exp{\left(
                -\frac{x^2}{2\sigma^2} \right)}, 
\end{equation}
\noindent where $x$ is the distance to the center of the PSF, and
$\sigma$ the standard deviation of the PSF, linked to the full width
half maximum (FWHM) of the instrument:
\begin{equation}
            \sigma = \frac{\text{FWHM}}{2\sqrt{2\ln{2}}}.
\end{equation}
The convolution product is thus: 
\begin{equation}
            (I \ast g)(b) = \int_{-\infty}^{+\infty}I(b') \, g(b-b')
            \, \mathrm{d}b'. 
\end{equation}
In practice, the integral is limited to the range of $b$ plus or minus
3 times the standard deviation $\sigma$ of the gaussian kernel.
Outside of the range of $b$, the intensity is taken as a linear
extrapolation of the profile.

\section{Two-component model of the \mbox{C$^{18}$O 2--1} and \CII\,158\,$\mu$m
line emission in the Horsehead} 
\label{app:Appendix_CO_Cplus_toy}

In this Appendix, we present the results of applying the simple
two-component model to C$^{18}$O and C$^+$. These two species serve as
examples of line emission arising predominantly from the dense PDR and
from the extended PDR envelope, respectively. In contrast to H$_2$O,
the C$^{18}$O~2--1 and \CII\,158\,$\mu$m lines are
low–critical-density transitions, with $n_{\rm cr}$ of only a few
$\times$10$^{3}$\,cm$^{-3}$. This means that they are collisionally
excited \mbox{($T_{\rm ex}$\,$\simeq$\,$T_{\rm k}$)} in the envelope
and therefore do not scatter line emission originating from the dense
PDR, which we demonstrate here. These representative models are
designed to reproduce the observed line peak intensities and their
approximate spatial distribution qualitatively. Specifically,
C$^{18}$O~2--1 peaks toward the dense PDR, while
\mbox{\CII\,158\,$\mu$m} peaks further ahead. The physical conditions
and line-widths are the same as those of the H$_2$O models
(\mbox{Sect.~\ref{sec:scattering}}). Following the PDR model, the
adopted C$^{18}$O abundances relative to H nuclei are
$1.5 \times 10^{-7}$ in the dense PDR and $1 \times 10^{-8}$ in the
envelope. For C$^+$, we adopt 2$\times$10$^{-4}$ in the envelope and
2$\times$10$^{-6}$ in the dense PDR. These choices successfully
reproduce the observed intensity levels, indicating that the simple
two-component model is globally realistic.

Figure~\ref{fig:mtc_models_C18O_CII} shows the predicted radial
profiles of the \mbox{C$^{18}$O~2--1} and \mbox{\CII\,158\,$\mu$m}
line intensities for the dense PDR-only model (blue squares), the
envelope-only model (red squares), and the combined dense PDR plus
envelope model (black squares). This figure is analogous to
Fig.~\ref{fig:mtc_models_2} for the \mbox{$o$-H$_2$O\,557\,GHz} line.
As C$^{18}$O~2--1 and \mbox{\CII\,158\,$\mu$m} lines are collisionally
excited, the envelope does not produce scattering of line photons
emitted in the dense PDR (i.e., the red and black curves in the
envelope are the same). Furthermore, the C$^{18}$O~2--1 line is
optically thin, so the emission from the dense PDR is not
self-absorbed. For C$^+$, the abundance in the dense PDR is much lower
than toward the envelope, so any emission along the line of sight
originates from the foreground envelope.

\begin{figure}
  \centering 
  \includegraphics[scale=0.49, angle=0]{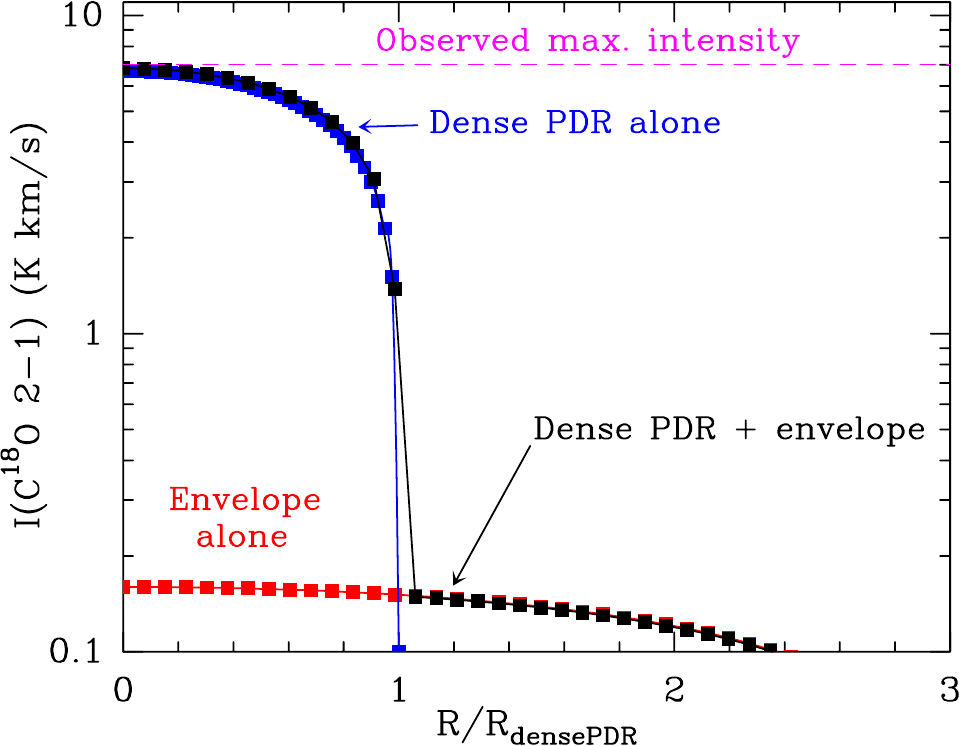}\vspace{0.5cm}
  \includegraphics[scale=0.49, angle=0]{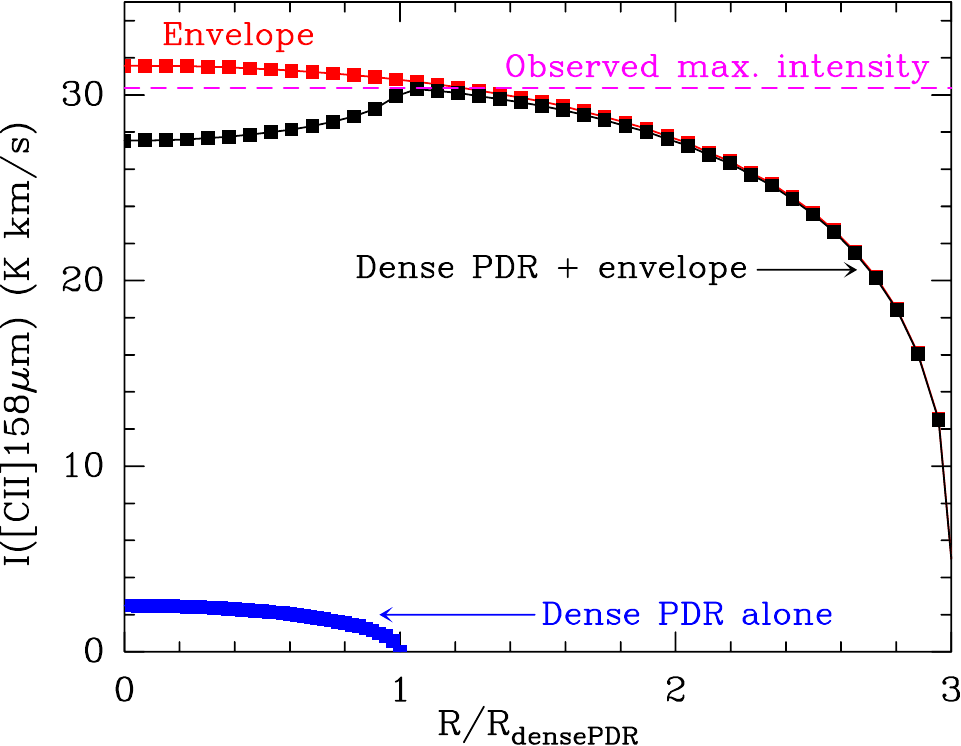}  
  \caption{Two-component, non-local non-LTE radiative transfer model of the Horsehead
  applied to C$^{18}$O and C$^+$.
  The horizontal magenta line shows the observed maximum line intensities
  of the C$^{18}$O~2--1 (upper panel) and \mbox{\CII\,158\,$\mu$m
    (lower panel) lines
    (Table~\ref{tab:peak}).}} \label{fig:mtc_models_C18O_CII} 
\end{figure}

\section{Weak dependence of the model results on \mbox{$G_{0}$}} 
\label{app:Appendix_Indep_to_G0}

We discuss here the dependence of the model line intensities on the
interstellar radiation field scaling factor, $G_{0}$. As shown below,
the best fit PDR wrapper model suggests a weak dependence of the model
line intensities on $G_0$, with the best fit $G_0$ value that appears
low compared to the nominal value of $G_0 \sim 100$ derived in
previous studies of the Horsehead nebula
\citep{Abergel2003,Habart2005,Goicoechea2009,SantaMaria2023}, and
implied by the spectral type of the ionizing star, $\sigma$~Ori, O 9.5
V at a projected distance of 3.5~pc. While observations suggest that
the ionizing star is located behind the nebula, the angle is quite
oblique (see Fig.~13 of \citealt{Abergel2024}). Therefore, the actual
distance is expected to be close to the projected distance.
\cite{Zannese2025} also favor $G_0 = 100$ in their analysis of new
JWST observations of H$_2$ line emission in the Horsehead nebula.
Analysis of observations of tens of H$_2$ lines obtained with the
IGRINS spectrometer on Gemini South (Piluso et al., in prep.) will
provide additional strong constraints on the value of $G_0$ in this
PDR. 

In principle, the difference in the $G_0$ values may be due to the new
modeling approach using the PDR wrapper or due to the particular
selection of tracers used in our analysis. To investigate the $G_0$
dependence of our model results, we first ran the minimization using
the standard 1D modeling approach, using the direct output intensities
of the plane-parallel isobaric Meudon PDR model grid. We verified that
the $\chi^2$ distribution is very flat in the parameter space
corresponding to the Horsehead physical conditions, and it favors low
pressure and $G_0$ values. We further note that \cite{Philipp2006}
also derived a low value of $G_0$ ($\sim 25$ in Draine units, or
$\sim 43$ in Habing units) from their isochoric models of the
Horsehead nebula based on the same CO isotopologue and [C\,{\sc i}]
observations as used here, but not including the newer SOFIA [C\,{\sc
  ii}] observations.

To further investigate the dependence of model results on $G_0$, we
constructed an extended grid of isobaric Meudon PDR models coupled
with the wrapper tool, explicitly including $G_{0}$ as a free
parameter. The explored parameter space covers:

\begin{itemize}
\item Thermal pressures $P_{\mathrm{th}} = 0.5$, 1, 2, 3, 4, 5, 6, 8,
  and $10 \times 10^6$~K cm$^{-3}$ ; 
\item Interstellar radiation fields $G_0 = 30$, 60, 70, 90, 100, and
  110 in Habing units ; 
\item Curvature radii $R_{\mathrm{C}} = 0.02$, 0.05, 0.1, 0.2, 0.5,
  and 1.0~pc. 
\end{itemize}

\begin{figure*}
\centering
\includegraphics[width=0.8\textwidth]{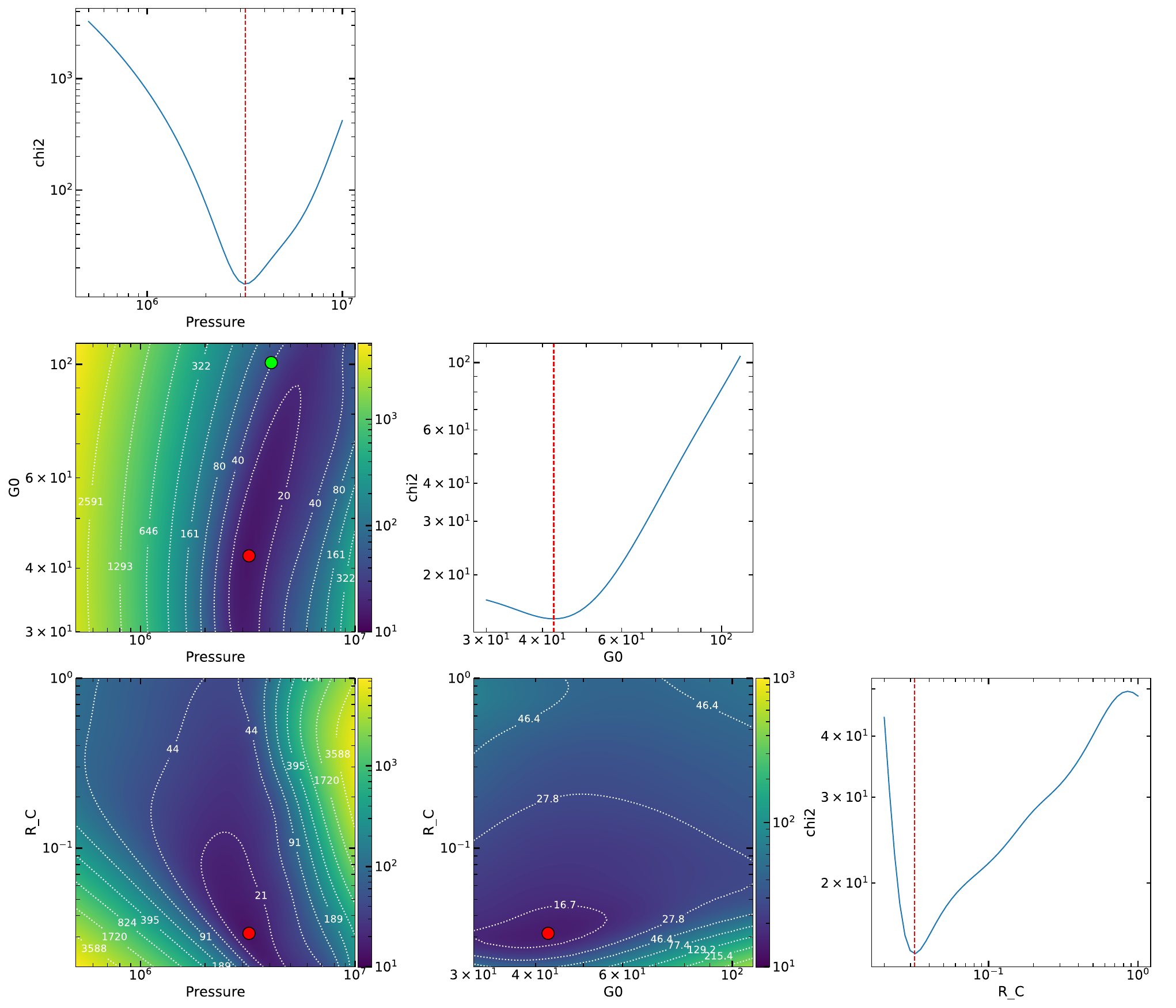} 
\caption{ $\chi^2$ maps obtained from the minimization procedure using
  maximum intensities and spatial separations. The best-fit solution,
  corresponding to $P_{\mathrm{th}} = 3.2 \times 10^6$~K\,cm$^{-3}$,
  $G_0 = 42$, and $R_{\mathrm{C}} = 0.032$~pc, is marked by a red
  circle. Constraints on $G_0$ and $R_{\mathrm{C}}$ remain weak, with
  the $\chi^2$ surface appearing relatively flat (bottom center
  panel), except for very small $R_{\mathrm{C}}$ values. For
  comparison, a model adopting literature values
  ($P_{\mathrm{th}} = 4 \times 10^6$~K\,cm$^{-3}$, $G_0 = 100$) is
  shown by a green circle in the $G_0$--$P_{\mathrm{th}}$ panel. }
\label{fig:app_chi2_maps}
\end{figure*}

\begin{figure*}
\centering
\includegraphics[width=0.85\textwidth]{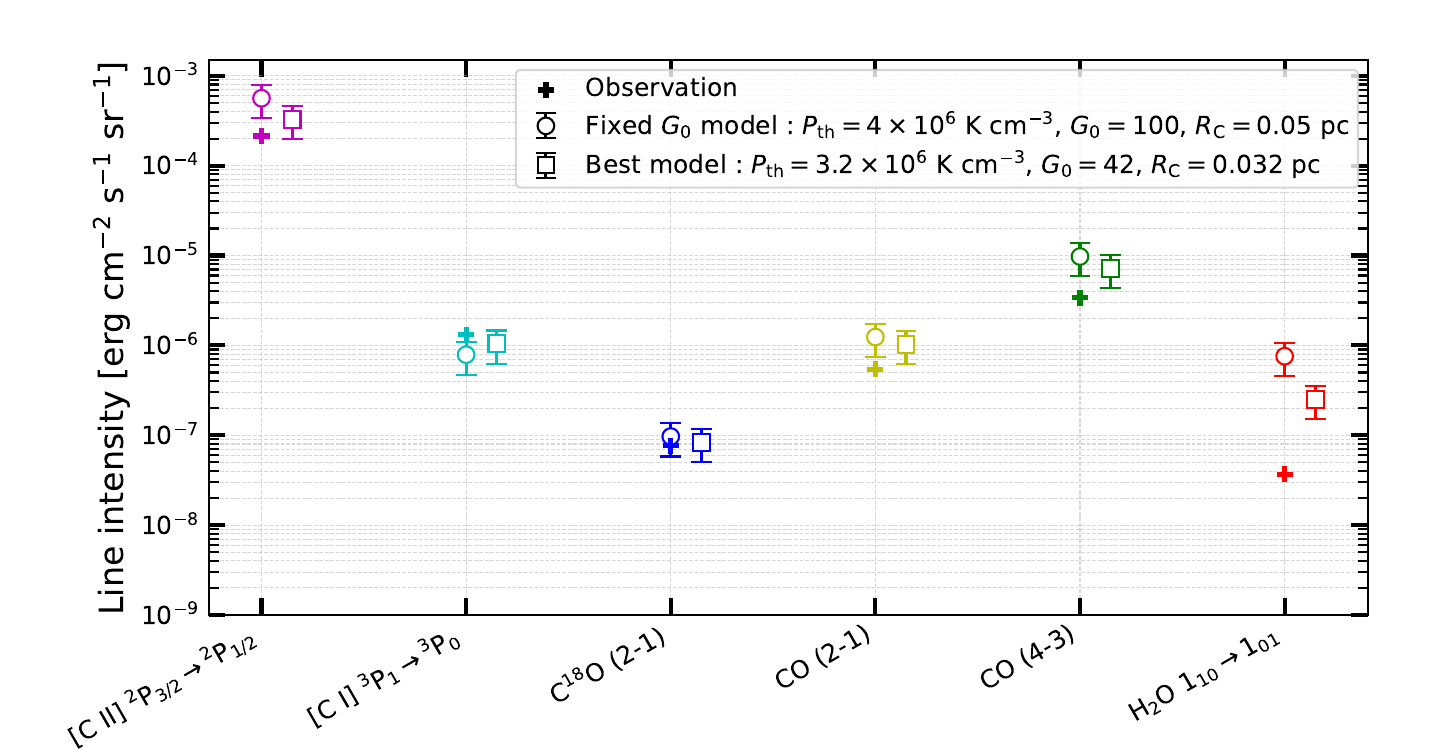} 
\caption{ Observed maximum intensities of various tracers (colored
  crosses) compared with model predictions (symbols with error bars).
  Squares show the best-fit model from the minimization described in
  Appendix~\ref{app:Appendix_Indep_to_G0}, while circles represent the
  model obtained in Section~\ref{sec:model} using literature
  parameters with $G_0 = 100$. All tracers are well reproduced, except
  for H$_2$O, for which the models over-predict the observed maximum
  intensity by about an order of magnitude. The error bars of 40\% on
  models represent geometrical uncertainties and the complexity of the
  parameter space. }
\label{fig:app_literature_model}
\end{figure*}

The resulting $\chi^2$ maps (Fig.~\ref{fig:app_chi2_maps}), obtained
from a minimization based on both line intensities and the spatial
separation of emission peaks along the PDR (see
Sections~\ref{sec:model_grid} and \ref{sec:model}), provide strong
constraints on the thermal pressure, which must lie between
$2.5 \times 10^6$ and $\sim 6 \times 10^6$~K cm$^{-3}$. By contrast,
the constraints on $G_0$ and $R_{\mathrm{C}}$ are much weaker: the
$\chi^2$ surface (bottom center panel) is relatively flat, with only a
mild preference for lower values. The formal minimum is found for
$P_{\mathrm{th}} = 3.2 \times 10^6$~K cm$^{-3}$, $G_0 = 42$, and
$R_{\mathrm{C}} = 0.032$~pc. However, as discussed above, such a low
$G_0$ value is inconsistent with previous studies of the Horsehead,
whereas $G_0 = 100$ is more consistent with the literature and remains
perfectly compatible with our constraints.

To illustrate the low sensitivity of $\chi^2$ to $G_0$, we show in
Fig.~\ref{fig:app_literature_model} an alternative model adopting
literature values for the parameters, most notably $G_0 = 100$. This
model provides a fit of comparable quality to the best-fit solution,
while also being more consistent with previous work, with
$P_{\mathrm{th}} = 4 \times 10^6$~K cm$^{-3}$, $G_0 = 100$, and
$R_{\mathrm{C}} = 0.05$~pc, i.e., about half the typical 0.1~pc width
usually assumed in plane-parallel models of the Horsehead PDR. The
best-fit model does improve the [C\,{\sc ii}]~158~$\mu$m intensity by
$\sim 30\%$, but this has no significant impact on the other tracers
(excluding H$_2$O, which was not included in the minimization).

The weak sensitivity to $G_0$ is expected. The intensities of
[C\,\textsc{ii}]~158~$\mu$m, [C\,\textsc{i}]~609~$\mu$m, and low-$J$
CO transitions vary only weakly with $G_{0}$ in the range
$30 \leq G_{0} \leq 100$, at thermal pressures typical of the
Horsehead ($P_{\rm th} \sim \text{a few } 10^{6}$~K cm$^{-3}$). This
behavior, already noted in earlier PDR studies \citep{Tielens1985a,
  Tielens1985b, Kaufman1999}, can be understood as follows: (i) in
Horsehead conditions, the [C\,\textsc{ii}] line is close to local
thermodynamical equilibrium and therefore saturated, so its intensity
is largely set by dust opacity and remains nearly constant with $G_0$;
(ii) the [C\,\textsc{i}] column density is itself almost insensitive
to the FUV field; and (iii) the low-$J$ CO lines are optically thick,
with intensities controlled by the temperature at the C$^{+}$/C/CO
transition, which varies little with $G_{0}$. As a result, the
$\chi^{2}$ maps display a nearly vertical structure in the
$(P_{\rm th}, G_{0})$ plane: the data constrain the thermal pressure
robustly, but provide little leverage on the incident FUV field.
Adopting $G_{0} = 100$, a value supported by previous studies of the
Horsehead, therefore does not affect our conclusions.

We further verified that models using the PDR wrapper with the
distance information about the stratification of the peak emission of
various tracers provide a tighter fit for the model parameters -- in
particular without the distance information, $G_0$ is much less
constrained. This suggests that observations at higher angular
resolution and including additional tracers would be required to
better constrain $G_0$.

The objective of the present manuscript is to study the water
vapor emission in the Horsehead nebula, not to determine the best
value of $G_0$. While models with lower $G_0$ values provide a
somewhat better fit to the water observations
(Fig.~\ref{fig:app_literature_model}), the model water line intensity
is stil a factor of 7 higher than the observations. Photon scattering
in a low-density warm envelope surrounding the dense PDR is thus still
required to explain our \emph{Herschel}/HIFI observations of the
557~GHz water line, regardless of the exact value of $G_0$ value used.

\end{document}